\documentclass[aps,prb,twocolumn,showpacs,10pt,floatfix,longbibliography]{revtex4-2}

\usepackage[english]{babel}
\usepackage{natbib}
\usepackage{amsfonts}
\usepackage{amsmath}
\usepackage{amssymb}
\usepackage{braket}
\usepackage{mathtools}
\usepackage{rotating}
\usepackage{color}
\usepackage{framed}
\usepackage{enumerate}
\usepackage{graphicx}
\usepackage{float}
\usepackage{hyperref}
\hypersetup{
    colorlinks=true,
    linkcolor=blue,    
    urlcolor=blue,
    citecolor=blue
    }
\usepackage{tikz}
\usepackage{circuitikz}
\usepackage{tabularx}
\usepackage{multirow}
\usepackage{csquotes}

\definecolor{blue-1}{rgb}{0.776,0.832,0.872}
\definecolor{blue-2}{rgb}{0.746,0.802,0.842}
\definecolor{blue-3}{rgb}{0.806,0.862,0.902}
\definecolor{gray-1}{rgb}{0.88,0.892,0.896}
\definecolor{gray-2}{rgb}{0.85,0.862,0.866}
\definecolor{gray-3}{rgb}{0.91,0.922,0.926}

\newcommand{\vm}{\vec m}
\newcommand{\vE}{\vec E}

\newcommand{\vi}{\vec i}
\newcommand{\vk}{\vec k}

\newcommand{\vis}{{\vec i}_{\rm s}}
\newcommand{\ve}{\vec e}

\newcommand{\vu}{{\vec u}}
\newcommand{\vus}{{\vec u}_{\rm s}}

\newcommand{\vrho}{\mbox{\boldmath $\rho$}}

\renewcommand{\vec}[1]{\mathbf{#1}}
\newcommand{\Dex}{D_{\rm ex}}
\newcommand{\zzeta}{f}
\newcommand{\calzzeta}{{\cal F}}
\newcommand{\ia}{j}
\newcommand{\ib}{k}

\begin{document}
\title{Magnon-mediated electric current drag and nonlocal spin-Peltier effect in the ac regime}

\author{Oliver Franke}
\author{Duje Akrap}
\author{Piet W. Brouwer}
\affiliation{Dahlem Center for Complex Quantum Systems and Physics Department, Freie Universit\"at Berlin, Arnimallee 14, 14195 Berlin, Germany}

\begin{abstract}
Electron-magnon coupling at the interface between a normal metal and a magnetically ordered insulator modifies the electrical conductivity of the normal metal, an effect known as spin-Hall magnetoresistance. It can also facilitate magnon-mediated electric current drag, the nonlocal electric current response of two normal metal layers separated by a magnetic insulator. Additionally, spin and heat transport are coupled both in the magnetic insulator and across the interfaces to normal metals. In this article, we present a theory of these spintronic and spin-caloritronic effects for time-dependent applied electric fields $E(\omega)$, with driving frequencies $\omega$ up to the THz regime. Our model describes how the dominant transport mechanism, coherent or incoherent magnons, evolves with the driving frequency $\omega$.
\end{abstract}

\maketitle

\newpage

\section{Introduction}
\label{sec:introduction}
In spintronic devices containing magnetic insulators a central role is played by the conversion between magnonic and electronic spin and heat currents at magnetic-insulator--normal-metal interfaces. Such conversion is at the heart of the spin-Seebeck effect \cite{Uchida2010-je, Bauer2012-tx}, the excitation of a spin current in a normal metal by a temperature gradient across the interface, and its inverse, the spin-Peltier effect. When combined with spin-charge conversion in the normal metal, which originates from the spin-Hall effect (SHE) and its inverse \cite{Dyakonov1971-pp, Hirsch1999-gc}, the interfacial magnon-electron coupling also enables key spintronic effects, such as current-induced magnetization switching \cite{Avci2017-kj,Yang2024-fm} and spin-Hall magnetoresistance (SMR) \cite{Weiler2012-gh,Huang2012-fu,Nakayama2013-gf,Hahn2013-rw,Vlietstra2013-uv,Althammer2013-zm,Lotze2014-qv,Choi2017-fn,Chen2013-gf,Chen2016-pc,Zhang2019-zv}.

In a geometry with multiple magnetic-insulator--normal-metal interfaces, nonlocal effects appear, in which electronic currents are converted into magnonic currents and back. An example of such an effect is magnon-mediated current drag. In this effect, an electrical current in one normal metal creates, via the spin-Hall effect and interfacial electronic-current-to-magnonic-current conversion, a magnonic spin current in an adjacent magnetic insulator. This magnonic spin current is then converted into an electronic spin current at the interface with a different normal metal, inducing a charge current in this second normal metal layer via the inverse spin-Hall effect \cite{Zhang2012-fy,Zhang2012-ig,Kajiwara2010-rj,Cornelissen2015-fh,Goennenwein2015-lb,Schlitz2021-ho,Li2016-ye,Wu2016-cs,Muduli2020-ub}. Such nonlocal spintronic effects not only provide information about the efficiency of interfacial conversion processes, unlike local spintronic effects, they also give valuable insights into magnonic relaxation channels in magnetic insulators.

\begin{figure}
  \centering
  \includegraphics[width=0.46\textwidth]{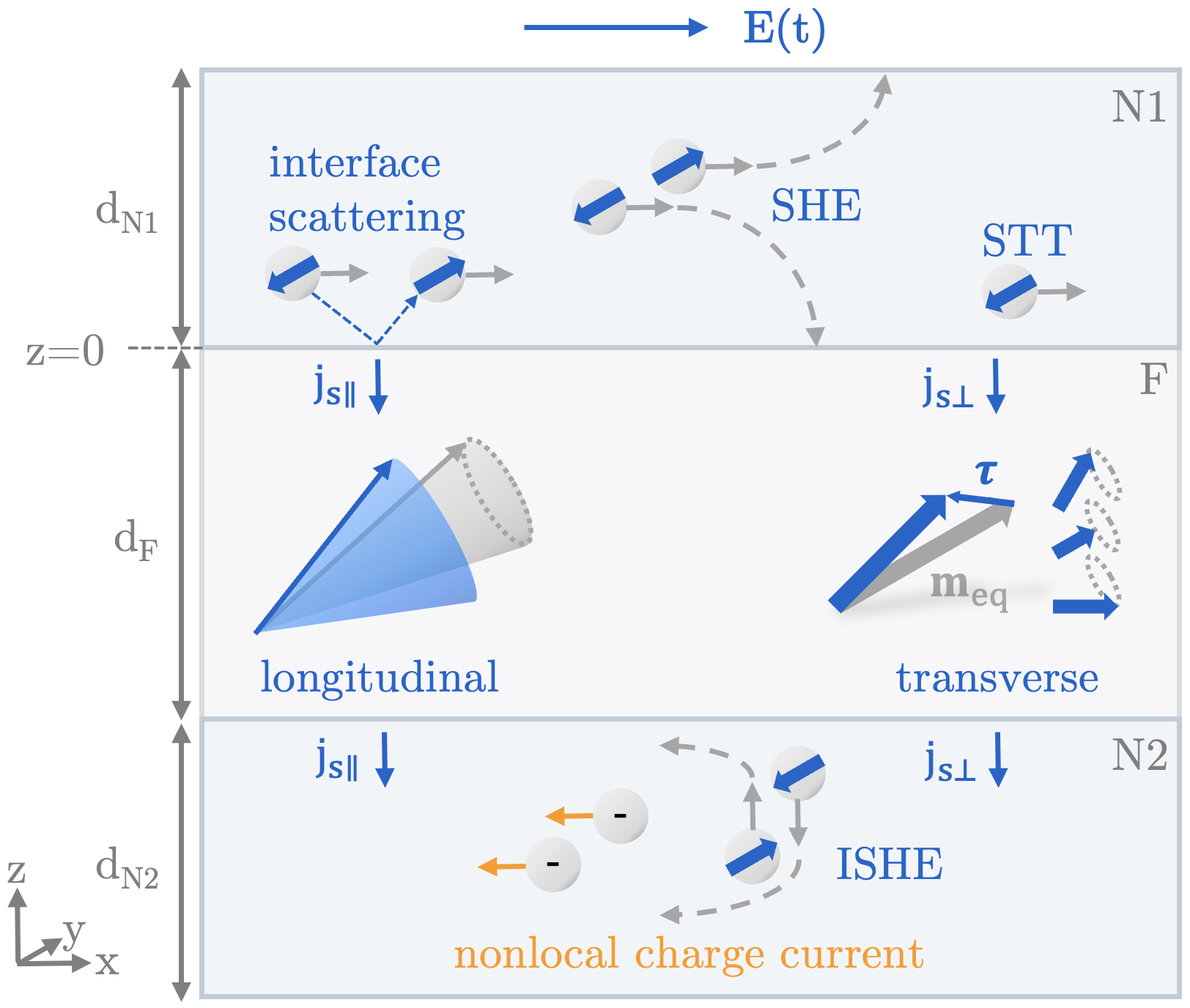}
\caption{Geometry of the N$|$F$|$N trilayer consisting of two normal metals N1 and N2 and a magnetically ordered insulator F. Via the spin-Hall effect (SHE), an in-plane electric field $\vE_1(t) = E(t) \ve_x$ in N1 causes a spin current through the F$|$N interfaces, which, via the inverse spin-Hall effect (ISHE), causes a correction to the charge current in N1 (not shown here). This correction to the charge conductivity is the spin-Hall magnetoresistance (SMR). In addition, a nonlocal current response arises in the N2 layer, which is referred to as magnon-mediated current drag. In the magnetic insulator F, the component $j_{{\rm s}\perp}$ of the spin current polarized perpendicular to the equilibrium magnetization direction $\vm_{\rm eq}$ is carried by coherent magnons, whereas the collinear component $j_{{\rm s}\parallel}$ is carried by incoherent, thermal magnons. The coherent magnons couple at the F$|$N interfaces via spin-transfer torque (STT) to the electronic spin accumulation in N. Incoherent magnons are excited by spin-flip scattering of electrons in N at the F$|$N interfaces. In this article, we calculate the charge currents $\vi_\ib$ linear in $E_\ia$ for driving fields $E_\ia(t) \propto \cos(\omega t)$ with driving frequencies ranging from the {\it dc} limit $\omega = 0$ to the THz regime.}
\label{fig:geometry}
\end{figure}

With the availability of experimental methods to investigate spintronic effects in the high-frequency THz regime \cite{Fulop2020-do,Walowski2016-zs}, it becomes necessary to also theoretically describe such effects in the limit of ultrafast driving. In this article and in the companion article \cite{Franke2025-nonlin}, we accomplish this task for spintronic effects in the prototypical geometry of an N$|$F$|$N trilayer (N: normal metal, F: ferromagnetic insulator), shown schematically in Fig.\ \ref{fig:geometry} and described in detail in Sec.~\ref{sec:linres}. The emphasis of the present article is on linear nonlocal transport in response to a time-dependent electric field $\vE_1(\omega)$ in one of the normal-metal layers. This includes magnon-mediated current drag --- a current $\vi_2(\omega)$ in the second N layer in response to the applied electric field $\vE_1(\omega)$ in the first --- and the nonlocal spin-Peltier effect, a temperature change $\delta T_2(\omega)$ linear in $\vE_1(\omega)$. The companion article \cite{Franke2025-nonlin} focuses on local and nonlocal nonlinear response to the applied electric field.

Our article builds on existing theories of linear spintronic effects in the {\it dc} limit \cite{Chen2013-gf,Chen2016-pc,Cornelissen2016-wy,Zhang2019-zv,Zhang2012-fy,Zhang2012-ig,Wang2018-qd} and extends them to the regime of high driving frequencies $\omega$. It also extends previous work on linear charge transport in F$|$N bilayers with {\it ac} driving by \citet{Reiss2021-em}.

Spin transport across the magnetic insulator F has contributions from coherent magnetization dynamics and from spin transport by incoherent (thermal) magnons. Which of these contributions dominates the response depends on the driving frequency $\omega$ and may be different for local and nonlocal spintronic effects. As a rule of thumb, coherent magnetization dynamics is relevant for the (local) spin-Hall magnetoresistance at low frequencies \cite{Chen2013-gf,Chen2016-pc} (although there also is a smaller contribution from thermal magnons \cite{Zhang2019-zv}) and at high frequencies near magnon resonances \cite{Sluka2017-xa,Johansen2018-az}, whereas incoherent magnons dominate the response otherwise \cite{Reiss2021-em}. On the other hand, the nonlocal response is mediated by incoherent magnons at low frequencies \cite{Zhang2012-fy,Zhang2012-ig}, whereas coherent magnetization dynamics is relevant near resonances only. The theory presented in this article must therefore address spin transport by coherent and incoherent magnons on equal footing, whereby coherent and incoherent magnons constitute parallel channels of spin transport through F.

For the description of nonlocal response across multiple interfaces, we find it advantageous to use an effective magneto-electric circuit theory \cite{Reiss2021-em}. In this approach, the response of each individual part of the system --- the two normal-metal layers N1 and N2, the magnetic layer F, and the two interfaces N1$|$F and F$|$N2 --- is described with the help of an effective impedance. In a second step, the response of the total system is found using Kirchhoff-like laws to combine these impedances. The impedances relate spin and heat currents to their ``driving forces'', which are spin accumulation and temperature for N1 and N2 and magnon chemical potential and temperature for F. Magnonic relaxation processes, which are essential for the nonlocal response across the magnetic insulator layer, are included in the impedance of the F layer. This circuit-based approach unfolds its full potential when applied not only to linear response but also to nonlinear phenomena --- an endeavor we undertake in our companion article, Ref.~\onlinecite{Franke2025-nonlin}.

The frequency dependence of spintronic effects in an N$|$F$|$N trilayer has its origin in the frequency dependence of magnon transport through the magnetic insulator F \cite{Reiss2021-em}. For incoherent magnons, relevant time scales are the various elastic and inelastic relaxation times and the time required to diffuse across F \cite{Cornelissen2016-wy,Schmidt2021-qo}. For coherent magnons, the relevant time scale is the time-of-flight between the two F$|$N interfaces. For YIG, these time scales result in characteristic frequencies in the THz range and below, depending on device parameters. On the other hand, the fundamental processes coupling spin and charge currents at the two F$|$N interfaces, as well as the spin-charge conversion in N mediated by the spin-Hall effect and its inverse, may be considered frequency-independent, at least for frequencies up to the THz range. The reason is that, unless the interface is very disordered, the transit time of electrons and magnons through the interface region is simply too short to result in an appreciable frequency dependence. Indeed, the spin-transfer torque (STT) \cite{Schellekens2014-se,Razdolski2017-st}, the interfacial spin-Seebeck effect \cite{Kimling2017-qd,Seifert2018-vz}, and the spin-Hall effect and its inverse \cite{Kampfrath2013-ke,Seifert2018-vj} have been shown to persist unchanged well into the THz frequency range.

In this article, we estimate the magnitude of the current response for a prototypical Pt$|$YIG$|$Pt trilayer. For this material combination, we find that up to driving frequencies in the low GHz range the SMR effect and the magnon-mediated current drag effect are of comparable magnitude to the {\it dc} limit, which has been observed experimentally \cite{Weiler2012-gh,Huang2012-fu,Nakayama2013-gf,Hahn2013-rw,Althammer2013-zm,Lotze2014-qv,Wu2016-cs}. For higher frequencies, up to the THz regime, the nonlocal signal is suppressed, except at magnon resonances where its strength may exceed the {\it dc} limit. For all frequencies considered, the nonlocal magnon-mediated current drag effect is at most as efficient as the local SMR effect, which typically gives a correction to the charge conductivity of order $\sim 10^{-4}$ \cite{Weiler2012-gh,Huang2012-fu,Nakayama2013-gf,Hahn2013-rw,Althammer2013-zm,Lotze2014-qv}.

The remainder of this article is organized as follows: We review the system geometry and introduce the basic elements of the magneto-electric circuit theory in Sec.\ \ref{sec:linearresponse}. In Sec.\ \ref{sec:conductivities} we combine the spin impedances defined in Sec.\ \ref{sec:linearresponse} to obtain the nonlocal electrical current response to an applied electric field in one of the normal-metal layers. This is the magnon-mediated current drag effect. In Sec.\ \ref{sec:peltier} we consider the temperature change linear in the applied electric field. This is the spin-Peltier effect. We conclude in Sec.\ \ref{sec:discussion}. The appendices contain details of our calculation. Our findings are illustrated by a numerical evaluation of local and nonlocal conductivities for a Pt$|$YIG$|$Pt trilayer. To facilitate the evaluation of our results for device parameters not considered by us or for other material combinations, an open source code is available to evaluate the local and nonlocal response for different materials and system sizes \cite{Franke2025-zenodo}.

\section{Linear response and magneto-electric circuit elements}
\label{sec:linearresponse}

We first review the fundamental linear-response relations between currents and their driving forces for the separate parts of the N$|$F$|$N trilayer. The currents are spin, charge, and heat currents. The driving forces are the applied electric field, the spin accumulation in the normal metals, the magnon chemical potential, and the electron and magnon temperatures. In Secs.\ \ref{sec:normalmetal}, \ref{sec:interfaces}, and \ref{sec:ferromagnet}, we present separate relations relating currents to the driving forces for the bulk of the two normal-metal layers N1 and N2, across the F$|$N interfaces, and for the bulk of the ferromagnetic insulator F, respectively. Following Ref.~\onlinecite{Reiss2021-em}, we formulate these relations in terms of impedances, the coefficients of proportionality between current and its driving force. Before that, in Sec.\ \ref{sec:linres}, we review some essential aspects of the setup we consider.

\subsection{System geometry and notation}
\label{sec:linres}

The N$|$F$|$N trilayer we consider is shown schematically in Fig.\ \ref{fig:geometry}. It consists of two normal metals N1 and N2, separated by a ferromagnetic insulator F. 
We choose coordinates such that N1 and N2 are located at $0 < z < d_{{\rm N}1}$ and $-d_{\rm F} - d_{{\rm N}2} < z < -d_{\rm F}$, whereas the ferromagnetic insulator F is at $-d_{\rm F} < z < 0$. Spatially uniform time-dependent electric fields $\vE_\ia(t) = E_\ia(t) \ve_x$, $\ia=1,2$, are applied in N1 and N2 in the $x$-direction.

In the following, we denote spatial directions by superscripts and spin directions by subscripts or boldface vector notation.
The (position dependent) magnetization direction of the ferromagnetic insulator F is denoted by the unit vector $\vm(z,t)$, where $-d_{\rm F} < z < 0$. The equilibrium magnetization direction is written as
\begin{equation}
    \mathbf{m}_{\rm eq} = 
  m_x \mathbf{e}_x + m_y \mathbf{e}_y + m_z \mathbf{e}_z.
\label{eq:equilibriummagnetization}
\end{equation}
We will find it useful to decompose vector-valued variables, such as the magnetization direction $\vm(z,t)$, the spin current $\vis^z(z,t)$, or the spin accumulation $e \vus(z,t)$ into components parallel to and perpendicular to $\vm_{\rm eq}$. Hereto, we choose a complex unit vector $\mathbf{e}_\perp$, whose real and imaginary parts span the perpendicular plane to $\vm_{\rm eq}$ and that fulfills
\begin{equation}
    \mathbf{e}_\perp \times \vm_{\rm eq} = i \mathbf{e}_\perp.
\label{eq:basiscrossproduct}
\end{equation}
A convenient choice is \cite{Reiss2021-em}
\begin{align}
\begin{split}
    \mathbf{e}_\perp = &\ \frac{1}{\sqrt{2(m_x^2+m_z^2)}} \big[(m_x^2+m_z^2) \mathbf{e}_y \\
    &\, - (m_z m_y - i m_x) \mathbf{e}_z - (m_x m_y + i m_z) \mathbf{e}_x \big].
\end{split}
\label{eq:eperp}
\end{align}
Writing the magnetization $\vm(z,t)$ in the basis $\{ \mathbf{m}_{\rm eq}, \mathbf{e}_\perp, \mathbf{e}_\perp^* \}$, we have
\begin{equation}
  \vm(z,t) = m_{\parallel}(z,t) \vm_{\rm eq}
  + m_{\perp}(z,t) \ve_{\perp} + m_{\perp}^*(z,t) \ve_{\perp}^*,
  \label{eq:magbasis}
\end{equation}
where $m_{\parallel}(z,t) = \sqrt{1 - 2 |m_{\perp}(z,t)|^2}$ because the vector $\vm(z,t)$ is normalized. The decomposition into components parallel to and perpendicular to $\vm_{\rm eq}$ for other vector-valued variables are analogous.
Because the dynamical variables in N1, N2, and F are mainly needed at the ferromagnet--normal-metal interfaces at $z=0$ and $z=-d_{\rm F}$, we use the short-hand notation $\vm_{1}(t) = \vm(0,t)$, $\vm_{2}(t) = \vm(-d_{\rm F},t)$ and analogously for the other variables.

Since the applied electric fields $E_\ia(t)$ are real, its Fourier transform $E_\ia(\omega)$ obeys $E_\ia(-\omega) = E^*_\ia(\omega)$. The transverse component of a vector-valued variable, such as $m_{\perp}(z,t)$, is complex, so that its Fourier transforms at frequencies $\omega$ and $-\omega$ are not complex conjugates of each other.

The goal of the calculations in this article is to calculate the charge currents $i^{x/y}_\ib(z,\omega)$ and change in temperatures $\delta T_\ib(z,\omega)$ in N1 and N2 to linear order in the applied electric fields. Our result for the charge current response is expressed in the form of local ($\ia=\ib$) and nonlocal ($\ia \neq \ib$) charge conductivities $\sigma_{\ib\ia}(\omega)$, that relate the applied field $E_\ia(\omega)$ in normal metal layer $\ia=1,2$ to the spatially averaged charge current density $\bar i^{x/y}_\ib(\omega)$ in layer $\ib=1,2$,
\begin{align}
  \bar i_\ib^x(\omega) &= \sum_{\ia=1}^{2} \sigma^{xx}_{\ib\ia}(\omega) E_\ia(\omega),
  \label{eq:linear_conductivity_definition_x} \\
  \bar i_\ib^y(\omega) &= \sum_{\ia=1}^{2} \sigma^{yx}_{\ib\ia}(\omega) E_\ia(\omega).
\label{eq:linear_conductivity_definition_y}
\end{align}
Our result for the spin-Peltier effect is a linear relation of the form
\begin{equation}
  \overline{\Delta T}_{{\rm e}\ib }(\omega) = \sum_{\ia=1}^{2} \eta^{x}_{\ib\ia}(\omega) E_\ia(\omega),
  \label{eq:Peltiereffect}
\end{equation}
where $\overline{\Delta T}_{{\rm e}\ia}(\omega)$ is the spatially averaged change of the electron temperature in N$\ia$, $\ia=1,2$. (The response coefficients $\eta^{x}_{\ib\ia}(\omega)$ are related to the Peltier coefficient, as discussed in Sec.\ \ref{sec:peltier}.) Results for the nonlocal conductivities $\sigma^{x/yx}_{\ib\ia}(\omega)$ and the response coefficients $\eta^{x}_{\ib\ia}(\omega)$ are given in Secs.\ \ref{sec:conductivities} and \ref{sec:peltier}, respectively. 

\subsection{Normal metal}
\label{sec:normalmetal}

Transport of charge and spin in the normal metals N1 and N2 is coupled via the spin-Hall effect and the inverse spin-Hall effect. To linear order in the applied field and the induced potential gradients, the charge current densities $i^{x,y}(z,\omega)$ and the spin current density $\vis^z(z,\omega)$ satisfy the phenomenological response equations \cite{Dyakonov1971-pp,Dyakonov1971-yh,Hirsch1999-gc,Takahashi2006-tz}
\begin{align}
  i^x (z, \omega) =&\ \sigma_{{\rm N}\ia} E_\ia(\omega) - \theta_{\text{SH}\ia} \frac{\sigma_{{\rm N}\ia}}{2} \frac{\partial}{\partial z} u_{{\rm s}y} (z, \omega) , \label{eq:ix} \\
  i^y (z, \omega) =&\ \theta_{\text{SH}\ia} \frac{\sigma_{{\rm N}\ia}}{2} \frac{\partial}{\partial z} u_{{\rm s}x} (z, \omega), \label{eq:iy} \\
  \vis^z (z, \omega) =&\, - \frac{\sigma_{{\rm N}\ia}}{2} \frac{\partial}{\partial z} \vus (z, \omega) - \theta_{\text{SH}\ia} \sigma_{{\rm N}\ia} E_\ia(\omega) \ve_{y},
  \label{eq:jsz}
\end{align}
where we write the indices $\ia=1,2$ in accordance with the position $z$ in N1 or N2. Here, $\theta_{{\rm SH}\ia}$ is the spin-Hall angle and $\sigma_{{\rm N}\ia}$ the electrical conductivity of layer N$\ia$. The spin current $\vis^z(z,\omega)$ is defined in units of an equivalent charge current, such that $i_{{\rm s}z}^z = i_{\uparrow}^z - i_{\downarrow}^z$, with $i_{\sigma}^z$ the charge current density carried by electrons with spin $\sigma$ with respect to the $z$ axis, and analogous for $i^z_{{\rm s}x}$ and $i^{z}_{{\rm s}y}$. Similarly, the spin accumulation is defined in terms of equivalent voltage units, $e u_{{\rm s}z} = \mu_{\uparrow} - \mu_{\downarrow}$, with $\mu_{\sigma}$ the chemical potential for electrons with spin $\sigma$ with respect to the $z$ axis, and analogous for $u_{{\rm s}x}$ and $u_{{\rm s}y}$.

The spin accumulation also determines the spin density $\vrho_{{\rm es}}(z,\omega)$,
\begin{equation}
  \vrho_{{\rm es}}(z,\omega) = e^2 \nu_{{\rm N}\ia} \vu_{{\rm s}}(z,\omega),
\label{eq:spindensity}
\end{equation}
where $\nu_{{\rm N}\ia}$ is the electronic density of states in N$\ia$, $\ia=1,2$, and the spin density is measured in equivalent charge units. Spin current and spin density satisfy the continuity equation
\begin{equation}
  -i \omega \vrho_{{\rm es}}(z, \omega)
  + \frac{\partial}{\partial z} \vis^z(z,\omega)
  = - g_{{\rm e}\mu,\ia} \vu_{{\rm s}}(z,\omega),
  \label{eq:spincontinuity}
\end{equation}
where
\begin{equation}
  g_{{\rm e}\mu,\ia} = 
  e^2 \frac{\nu_{{\rm N}\ia}}{\tau_{{\rm sf},\ia }},
  \label{eq:gspincontinuity}
\end{equation}
with $\tau_{{\rm sf},\ia }$ the spin flip time in N$\ia$. Combining Eqs.~(\ref{eq:jsz})--(\ref{eq:gspincontinuity}), one finds that the spin accumulation satisfies
\begin{equation}
  \frac{\partial^2}{\partial z^2} \vus(z,\omega) =
  \frac{1}{\lambda_{{\rm N}\ia}(\omega)^2} \vus(z,\omega),
\end{equation}
where
\begin{equation}
  \lambda_{{\rm N}\ia}(\omega)^2 = \frac{\sigma_{{\rm N}\ia}}{2 e^2 \nu_{{\rm N}\ia}(1/\tau_{{\rm sf},\ia } - i \omega)}
  \label{eq:lambdaNi}
\end{equation}
is the spin relaxation length in N$\ia$, $\ia=1,2$. We will only consider frequencies $\omega \ll 1/\tau_{{\rm sf},\ia }$ and, hence, neglect the frequency dependence of $\lambda_{{\rm N}\ia}$ throughout. This is an excellent approximation for Pt, which is the metal we choose for our numerical evaluation in Sec.\ \ref{sec:numericalestimates}. For metals with larger spin relaxation rates, such as W, for which $1/\tau_{\rm sf}$ is the THz range \cite{Nair2021-kk}, one simply replaces $\lambda_{{\rm N}\ia} \to \lambda_{{\rm N}\ia}(\omega)$ in the following Eqs.~\eqref{eq:zn} and \eqref{eq:linearsourceterm} to accurately describe the spin accumulation dynamics \cite{Vedyaev2020-ss}.

We assume that $\lambda_{{\rm N}\ia}$ is much smaller than the thickness $d_{{\rm N}\ia}$ of each normal metal layer, so that only the spin accumulations $\vu_{{\rm s}\ia}(\omega)$ at the interfaces between F and N$\ia$ at $z=0$ (for $\ia=1$) and $z=-d_{\rm F}$ (for $\ia=2$) contribute to the inverse spin-Hall effect. From Eqs.~(\ref{eq:ix}) and (\ref{eq:iy}) one then finds that the spatially averaged corrections to the charge current densities in N1 and N2 read
\begin{align}
  \label{eq:deltaix}
  \delta \bar i_\ia^{x}(\omega) =&\ (-1)^{\ia-1} \theta_{{\rm SH}\ia}
     \frac{  \sigma_{{\rm N}\ia}}{2 d_{{\rm N}\ia}} 
     u_{{\rm s}\ia y}(\omega), \\
  \label{eq:deltaiy}
  \delta \bar i_\ia^{y}(\omega) =&\, -(-1)^{\ia-1} \theta_{{\rm SH}\ia }
     \frac{\sigma_{{\rm N}\ia }}{2 d_{{\rm N}\ia }} 
    u_{{\rm s}\ia x}(\omega),\ \ \ia=1,2,
\end{align}
where $e u_{{\rm s}\ia y/x}(\omega)$ is the spin accumulation at the interface between N$\ia$ and F,
whereas Eq.~\eqref{eq:jsz} gives a relationship between the spin voltage $\vu_{{\rm s}\ia }(\omega)$ and the spin current $\vi_{{\rm s}\ia }(\omega)$ at the interface,
\begin{align}
  \label{eq:nmspinrelation}
  (-1)^{\ia-1} Z_{{\rm N}\ia }(\omega) \vi_{{\rm s}\ia }(\omega) =
  \vu_{{\rm s}\ia }(\omega) - \delta \vu_{{\rm s}\ia }(\omega).
\end{align}
Here, the ``spin impedance'' $Z_{{\rm N}\ia }$, $\ia=1,2$, is \cite{Reiss2021-em}
\begin{equation}
  Z_{{\rm N}\ia } = \frac{2 \lambda_{{\rm N}\ia }}{\sigma_{{\rm N}\ia }}
\label{eq:zn}
\end{equation}
and $\delta \vu_{{\rm s}\ia }(\omega)$ a source term proportional to the applied electric field,
\begin{equation}
  \delta \vu_{{\rm s}\ia }(\omega) = 2 (-1)^{\ia-1}
  \lambda_{{\rm N}\ia } \theta_{{\rm SH}\ia } E_\ia (\omega) \ve_y.
\label{eq:linearsourceterm}
\end{equation}

At the ferromagnet--normal-metal interfaces and in the bulk ferromagnet, spin and heat transport is coupled (see Secs.~\ref{sec:interfaces} and \ref{sec:ferromagnet}). To ensure that response relations for spin and heat have the same form, we use units, such that heat current density $(k_{\rm B} T/2 e) i^z_{\rm Q}(z,t)$, energy density $(k_{\rm B} T/2e) \rho_{{\rm eQ}}(z,t)$, and temperature change $\Delta T_{\rm e}(z,t) = (e/k_{\rm B}) u_{\rm eQ}(z,t)$ of the conduction electrons have the dimension of an equivalent charge current, charge density, and voltage, respectively. The temperature change $\Delta T_{\rm e}(z,t)$ is measured with respect to the lattice temperature $T$, which is assumed to remain constant throughout.
To describe heat transport in N1 and N2, we use the continuity equation for the energy density and heat current density
\begin{equation}
  -i \omega \rho_{{\rm eQ}}(z,\omega) +
  \frac{\partial}{\partial z} i^z_{\rm Q}(z,\omega)
  = - g_{{\rm eQ},\ia } u_{\rm eQ}(z,\omega),
\label{eq:heatcontinuity}
\end{equation}
where
$g_{{\rm eQ},\ia }$ is a constant describing the energy exchange with the lattice,
\begin{equation}
  g_{{\rm eQ},\ia } = \frac{2 e^2}{k_{\rm B}^2 T} \frac{C_{{\rm e}\ia }}{\tau_{{\rm ep},\ia }}.
\label{eq:gheatcontinuity}
\end{equation}
Here, $C_{{\rm e}\ia }$ is the electronic heat capacity and $\tau_{{\rm ep},\ia }$ the characteristic electron-phonon relaxation time in N$\ia$, $\ia=1,2$. 
The energy density $\rho_{\rm eQ}(z,\omega)$ is related to the electron temperature, while the heat current density $i^z_{\rm Q}(z,\omega)$ is proportional to its gradient,
\begin{align}
  \rho_{\rm eQ}(z,\omega) =&\ \frac{2 e^2}{k_{\rm B}^2 T} C_{{\rm e}\ia } u_{\rm eQ}(z,\omega),  \\
  i^z_{\rm Q}(z,\omega) =&\, - \frac{2 e^2}{k_{\rm B}^2 T} \kappa_{{\rm e}\ia } \frac{\partial}{\partial z} u_{\rm eQ}(z,\omega),
  \label{eq:jqz}
\end{align}
with $\kappa_{{\rm e}\ia }$ the electronic heat conductivity in N$\ia$, $\ia=1,2$. Equation (\ref{eq:heatcontinuity}) then becomes
\begin{equation}
  \frac{\partial^2}{\partial z^2}
  u_{\rm eQ}(z,\omega) =
  \frac{1}{l_{{\rm ep},\ia }(\omega)^2}
  u_{\rm eQ}(z,\omega),
\label{eq:nmheatdiffusion}
\end{equation}
 with the thermal relaxation length
\begin{align}
  l_{{\rm ep},\ia }(\omega)^2 &= \frac{\kappa_{{\rm e}\ia }}{C_{{\rm e}\ia }(1/\tau_{{\rm ep},\ia } -i \omega)}, \ \ \ia=1,2.
  \label{eq:thermalrelaxationlength}
\end{align}
With the boundary conditions that the heat currents vanish at the interfaces with vacuum at $z=d_{{\rm N}1}$ and $z=-d_{{\rm N}2} - d_{\rm F}$, we find that the excess electron temperature $u_{{\rm eQ}\ia }$ and the electronic heat current $i_{{\rm Q}\ia }$ at the two ferromagnet--normal-metal interfaces $\ia=1,2$ are related as
\begin{equation}
  (-1)^{\ia-1} Z_{{\rm QN}\ia }(\omega) i_{{\rm Q}\ia }(\omega) = u_{{\rm eQ}\ia } (\omega) - \delta u_{{\rm eQ}\ia }(\omega),
\label{eq:nmheatrelation}
\end{equation}
where
\begin{equation}
  Z_{{\rm QN}\ia }(\omega) = \frac{k_{\rm B}^2 T}{2 e^2} \frac{l_{{\rm ep},\ia }(\omega)}{\kappa_{{\rm e}\ia }}
  \coth \frac{d_{{\rm N}\ia }}{l_{{\rm ep},\ia }(\omega)}
\label{eq:zqn}
\end{equation}
is a ``thermal impedance'' and $\delta u_{{\rm eQ}\ia }(\omega)$ a source term, which is included for future reference. (The source term $\delta u_{{\rm eQ}\ia }(\omega)$ represents the effect of local heating. It must be set to zero when considering the linear response to an applied electric field.)

Anticipating that heat and longitudinal spin current are coupled at the F$|$N interfaces and inside F, we combine the longitudinal component (with respect to the equilibrium magnetization direction, see Eq.~\eqref{eq:magbasis}) of the spin current, $i_{{\rm s}\ia \parallel}(t)$, and the heat current $i_{{\rm Q}\ia }(t)$ into a single two-component vector,
\begin{equation}
  {\cal I}_\ia(\omega) = \begin{pmatrix} i_{{\rm s}\ia \parallel}(\omega) 
  \\ i_{{\rm Q}\ia }(\omega) \end{pmatrix},
  \label{eq:calI}
\end{equation}
with similar definitions for ${\cal U}_{{\rm e}\ia }(\omega) = (u_{{\rm s}\ia  \parallel}(\omega),u_{{\rm eQ}\ia }(\omega))^{\rm T}$ and $\delta {\cal U}_{{\rm e}\ia }(\omega) = (\delta u_{{\rm s}\ia  \parallel}(\omega),\delta u_{{\rm eQ}\ia }(\omega))^{\rm T}$. Consequently, we also combine the impedances $Z_{{\rm N}\ia }$ and $Z_{{\rm QN}\ia }$ for spin and heat transport into a $2 \times 2$ matrix
\begin{equation}
  {\cal Z}_{{\rm N}\ia }(\omega) = \begin{pmatrix} Z_{{\rm N}\ia } & 0 \\ 0 & Z_{{\rm QN}\ia }(\omega) \end{pmatrix}.
\label{eq:znmat}
\end{equation}

\subsection{F$|$N interfaces}
\label{sec:interfaces}

At the fundamental level, the coupling between magnonic and electronic spin and heat currents at the interface between a magnetic insulator and a normal metal has its origin in the spin-transfer torque \cite{Berger1996-xi,Slonczewski1996-is} and spin pumping \cite{Tserkovnyak2002-ax}.
For both effects, the strength of the coupling across the interface is described by the spin-mixing conductance \cite{Brataas2000-ar}.
Whereas the spin-transfer torque and spin pumping were originally discussed in the context of coherent magnetization dynamics, they are now understood to also govern the coupling of incoherent thermal magnons to electronic excitations in the normal metal \cite{Xiao2010-wi,Bender2015-tr}.

The coherent transverse component $i_{{\rm s}\ia \perp}(\omega)$ of the spin current across the ferromagnet--normal-metal F$|$N$\ia$ interface, $\ia=1,2$, depends on the transverse component $m_{\perp \ia}(\omega)$ of the magnetization and the transverse component $u_{{\rm s}\ia \perp}(\omega)$ of the spin accumulation in the normal metal. 
In linear response, the equation for the transverse components of the spin currents across the F$|$N interfaces are \cite{Tserkovnyak2002-ax,Tserkovnyak2002-hn}
\begin{equation}
  u_{{\rm s}\ia \perp}(\omega) +
  \frac{\hbar \omega}{e} m_{\perp \ia}(\omega) =
  -(-1)^{\ia-1} Z_{{\rm FN}\ia \perp} i_{{\rm s}\ia \perp}(\omega),
\label{eq:fnspinrelationperp}
\end{equation}
where $Z_{{\rm FN}\ia \perp}$ is the interfacial spin impedance \cite{Reiss2021-em}
\begin{align}
  Z_{{\rm FN}\ia \perp} =&\ \frac{1}{g_{\uparrow\downarrow \ia}},
  \label{eq:zfnperp}
\end{align}
with $g_{\uparrow\downarrow \ia}$ the spin-mixing conductance per unit area of the F$|$N$\ia$ interface, $\ia= 1,2$ \cite{Brataas2000-ar}.

The incoherent, longitudinal spin current through the interface depends not only on the magnon chemical potential $\mu_{{\rm m}\ia }$ and the longitudinal component $\mu_{{\rm s}\ia \parallel}$ of the spin accumulation at the interface \cite{Bender2015-tr,Reiss2022-tm}, but, via the interfacial spin-Seebeck effect, also on the difference $\Delta T_{{\rm e}\ia } - \Delta T_{{\rm m}\ia }$ of electron and magnon temperatures across it. At the same time, a temperature and/or spin accumulation/magnon chemical potential difference also causes a heat current through the interface \cite{Flipse2014-kh, Cornelissen2016-ti}. For an expression of the longitudinal spin current and the heat current at the two interfaces we again switch to equivalent charge units, writing $u_{{\rm m}\ia } = \mu_{\rm m}/e$ and $u_{{\rm mQ}\ia } = k_{\rm B} \Delta T_{{\rm m}\ia }/e$, and define the two-component spinor ${\cal U}_{{\rm m}\ia } = (u_{{\rm m}\ia }, u_{{\rm mQ}\ia })^{\rm T}$. One then has \cite{Reiss2021-em, Cornelissen2016-wy}
\begin{align}
  {\cal U}_{{\rm e}\ia } - {\cal U}_{{\rm m}\ia }
  =&\, - (-1)^{\ia-1} {\cal Z}_{{\rm FN}\ia \parallel}{\cal I}_\ia,
\label{eq:fnrelationpara}
\end{align}
where $\mathcal{Z}_{{\rm FN}\ia  \parallel}$ is the interfacial impedance matrix,
\begin{equation}
    \mathcal{Z}_{{\rm FN}\ia  \parallel}^{-1} = \frac{3 k_{\rm T}^3 \mathrm{Re} g_{\uparrow\downarrow \ia}}{16 \pi^{3/2} s} 
    \begin{pmatrix}
        4 \zeta(3/2) & 10 \zeta(5/2) \\
        10 \zeta(5/2) & 35 \zeta(7/2)
    \end{pmatrix}.
\label{eq:calzfnpara}
\end{equation}
Here, $k_{\rm T}$ is the thermal magnon wave number,
\begin{equation}
  k_{\rm T} = \sqrt{\frac{k_{\rm B} T}{\hbar D_{\rm ex}}}
\label{eq:thermalk}
\end{equation}
where $T$ is the equilibrium temperature, $s$ the equilibrium spin density in F, and $\zeta$ the Riemann zeta function. Note that all four elements of the $2 \times 2$ impedance matrix ${\cal Z}_{{\rm FN}\ia \parallel}$ have the same dimension, because spin and heat currents, as well as spin accumulation and temperature, are measured in the same units.

\subsection{Ferromagnetic insulator}
\label{sec:ferromagnet}

In the ferromagnetic insulator F, the spin current $\vi^z_{\rm s}(z,t)$ is carried by magnons. It has a coherent transverse component, which is related to the magnetization dynamics, as well as an incoherent longitudinal component, which is carried by thermal magnons. Below, we give expressions linking the magnonic spin and heat currents at the F$|$N interfaces to the two-component magnon potentials ${\cal U}_{{\rm m}1}(\omega)$ and ${\cal U}_{{\rm m}2}(\omega)$ and the magnetization amplitudes $m_{\perp 1}$ and $m_{\perp 2}$. For the thermal magnons, we employ a phenomenological description of magnon-mediated spin and heat transport, while the coherent magnetization dynamics is described by the Landau-Lifshitz-Gilbert equation.

\subsubsection{Coherent magnetization dynamics}
\label{sec:ferrocoherentmagnons}
The transverse magnetization amplitude $m_{\perp}(z,\omega)$ satisfies the linearized Landau-Lifshitz-Gilbert equation,
\begin{align}
  -\Dex
    \frac{\partial^2 }{\partial z^2} m_{\perp}(z,\omega)
  =&\,
  (\omega + i \alpha \omega - \omega_0) m_{\perp}(z,\omega),
  \label{eq:llglin}
\end{align}
where $\omega_0$ is the ferromagnetic-resonance frequency, which includes effects of static external magnetic fields, the demagnetization field, and anisotropies, $\Dex$ is the spin stiffness, and $\alpha$ the bulk Gilbert damping coefficient. The transverse spin current is given by
\begin{equation}
  j^z_{\rm s \perp}(z,\omega) =
  i \hbar \Dex s \frac{\partial}{\partial z} m_{\perp}(z,\omega).
  \label{eq:fijszperp}
\end{equation}
Solving Eq.~(\ref{eq:llglin}) with the boundary conditions $m_{\perp}(0,\omega) = m_{\perp 1}(\omega)$, $m_{\perp}(-d_{\rm F},\omega) = m_{\perp 2}(\omega)$, we find
\begin{align}
\begin{split}
  m_{\perp}(z,\omega) =&\
  \frac{m_{\perp 1}(\omega) \sin(k(\omega)(z+d_{\rm F}))}{\sin(k(\omega) d_{\rm F})} \\ &\,
  - \frac{m_{\perp 2}(\omega)
  \sin(k(\omega) z)}{\sin(k(\omega) d_{\rm F})},
\end{split}
\label{eq:mperp}
\end{align}
where $k(\omega)$ is the magnon wavenumber at frequency $\omega$,
\begin{equation}
  k(\omega)^2 = \frac{\omega(1+i\alpha) - \omega_0}{\Dex}.
  \label{eq:k}
\end{equation}
For definiteness, in expressions that contain $k(\omega)$, we choose the sign of $k(\omega)$ such that $\mbox{Im}\, k(\omega) > 0$.
Combining Eqs.~(\ref{eq:fijszperp}) and (\ref{eq:mperp}), one obtains a relation between the transverse spin currents and the transverse magnetization amplitudes at the two ferromagnet--normal-metal interfaces,
\begin{align}
\begin{split}
  &(-1)^{\ia-1} (\hbar \omega / e) \left[m_{\perp \ia}(\omega) \cos(k(\omega) d_{\rm F}) - m_{\perp \ib}(\omega) \right]  \\
  &\quad =\, 
    -i Z_{{\rm F}\perp}^{\infty}(\omega) i_{{\rm s}\ia  \perp}(\omega) \sin(k(\omega) d_{\rm F}), \ \ \ia \neq \ib,
\end{split}
\label{eq:fispinrelationperp}
\end{align}
where
\begin{equation}
  Z_{{\rm F}\perp}^{\infty}(\omega) = \frac{\hbar \omega}{2 e^2 \Dex k(\omega) s}.
\label{eq:zfperpinf}
\end{equation}

\subsubsection{Thermal magnons}
\label{sec:ferrothermalmagnons}

Following \citet{Cornelissen2016-wy} we describe thermal magnons in F in terms of the magnon chemical potential $\mu_{\rm m}(z,t)$ and the difference $\Delta T_{\rm m}(z,t) = T_{\rm m}(z,t) - T$ of the magnon temperature and the lattice temperature. We again switch to equivalent charge units and combine the magnon chemical potential $\mu_{\rm m}(z,t) = e u_{\rm m}(z,t)$ and temperature $\Delta T_{{\rm m}}(z,t) = e u_{\rm mQ}(z,t)/k_{\rm B}$ into the two-component spinor ${\cal U}_{{\rm m}}(z,t) = (u_{{\rm m}}, u_{{\rm mQ}})^{\rm T}$. We also combine the deviations of the magnon spin density $(\hbar/2e) \rho_{{\rm ms}}(z,t)$ and the magnon energy density $(k_{\rm B} T/2 e) \rho_{{\rm mQ}}(z,t)$, both measured in charge units, into the two-component spinor ${\cal P}_{\rm m}(z,t) = (\rho_{{\rm ms}}, \rho_{{\rm mQ}})^{\rm T}$. 

To linear order in ${\cal U}_{\rm m}(z,t)$, ${\cal P}_{\rm m}(z,t)$ may be expanded as
\begin{equation}
  \label{eq:drhoms}
  {\cal P}_{\rm m}(z,t)
  = \mathcal{C}_{\rm m}
  {\cal U}_{{\rm m}}(z,t),
\end{equation}
where the $2 \times 2$ matrix ${\cal C}_{\rm m}$ is a generalized heat capacity.

Expressions for the four elements of the matrix $\mathcal{C}_{\rm m}$ in terms of the magnon density of states are given in App.\ \ref{sec:boltzmann}. The magnon spin density and magnon energy density satisfy the continuity equation
\begin{align}
  \frac{\partial}{\partial t}
  {\cal P}_{\rm m}(z,t)
  + \frac{\partial}{\partial z}
  {\cal I}(z,t) = - \mathcal{G}_{\rm m}
  {\cal U}_{\rm m}(z,t),
\label{eq:diffeq_magnonheattransport}
\end{align}  
where ${\cal I}(z,t) = (i_{{\rm s}\parallel}^z(z,t),i_{\rm Q}^z(z,t))^{\rm T}$ is a two-component spinor combining the longitudinal magnon spin current and the magnon heat current, both measured in equivalent charge units. The $2 \times 2$ matrix ${\cal G}_{\rm m}$ describes the combined effect of various forms of spin and energy relaxation of magnons. In App.\ \ref{sec:boltzmann} we express ${\cal G}_{\rm m}$ in terms of the relaxation times $\tau_{\rm m,ex}$, $\tau_{\rm mp,ex}$, $\tau_{\rm m,rel}$, $\tau_{\rm mp,rel}$, and $\tau_{\rm el}$ for exchange-based spin-conserving magnon-magnon and magnon-phonon scattering, relativistic spin-non-conserving magnon-magnon and magnon-phonon scattering, and elastic magnon-impurity scattering.

Finally, the magnon spin and heat current densities depend linearly on gradients of the magnon chemical potential and the magnon temperature, 
\begin{equation}
    {\cal I}(z, \omega)
    = - \Sigma(\omega)
  \frac{\partial}{\partial z}
    {\cal U}_{{\rm m}}(z, \omega).
\label{eq:ferromagnetlinresponse}
\end{equation}
where $\Sigma_{\rm m}(\omega)$ is the generalized conductivity matrix \cite{Cornelissen2016-wy}
\begin{equation}
  \Sigma_{\rm m}(\omega) = \frac{1}{1 - i \omega \tau_{\rm m}}
    \begin{pmatrix}
        \sigma_{\rm m} & e L_{\rm m}/k_{\rm B} T \\
        e L_{\rm m}/k_{\rm B} T & 2 e^2 \kappa_{\rm m} / k_{\rm B}^2 T
    \end{pmatrix},\
    \label{eq:capitalsigma}
\end{equation}
with the magnon spin conductivity $\sigma_{\rm m}$, the bulk spin-Seebeck coefficient $L_{\rm m}$, and the magnon heat conductivity $\kappa_{\rm m}$, and
\begin{align}
  \tau_{\rm m}^{-1} =&\ \tau_{{\rm m},{\rm ex}}^{-1} + \tau_{{\rm m},{\rm rel}}^{-1} + \tau_{{\rm mp},{\rm ex}}^{-1} + \tau_{{\rm mp},{\rm rel}}^{-1} + \tau_{\rm m,el}^{-1}.
\label{eq:magnonlifetime}
\end{align}
Expressions relating $\sigma_{\rm m}$, $L_{\rm m}$, and $\kappa_{\rm m}$ to the relaxation time $\tau_{\rm m}$ are given in App.\ \ref{sec:boltzmann}. 

Combining Eqs.\ (\ref{eq:drhoms}), (\ref{eq:diffeq_magnonheattransport}), and (\ref{eq:ferromagnetlinresponse}), we find that the two-component spinor ${\cal U}_{\rm m}(z,t)$ obeys the diffusion-like equation
\begin{equation}
    \frac{\partial^2}{\partial z^2}
    {\cal U}_{{\rm m}}(z, \omega)
    = \Lambda(\omega)^{-2}
    {\cal U}_{{\rm m}} (z, \omega),
\label{eq:ferromagnetdiffusionequation}
\end{equation}
where
\begin{equation}
  \Lambda^2(\omega) = 
  (\mathcal{G}_{\rm m} - i \omega \mathcal{C}_{\rm m})^{-1}
  \Sigma(\omega).
\label{eq:capitallambda}
\end{equation}
We choose the sign of $\Lambda(\omega)$ such that $\mbox{Re}\, l_{m}(\omega) > 0$ for the two eigenvalues $l_{m}(\omega)$ of $\Lambda(\omega)$, $m=1,2$.
The solution of Eq.~\eqref{eq:ferromagnetdiffusionequation} can then be written in a form similar to Eq.~(\ref{eq:mperp}),
\begin{align}
\begin{split}
    {\cal U}_{{\rm m}}(z, \omega) =&\
    \frac{\sinh(\Lambda(\omega)^{-1}(z + d_{\rm F}))}{\sinh(\Lambda(\omega)^{-1} d_{\rm F})}
    {\cal U}_{{\rm m}1}(\omega) \\ &\,
    - \frac{\sinh{(\Lambda(\omega)^{-1} z)}}{\sinh{(\Lambda(\omega)^{-1} d_{\rm F})}}
    {\cal U}_{{\rm m}2}(\omega),
\end{split}
\label{eq:ferromagnetdiffsolution}
\end{align}
where ${\cal U}_{{\rm m}1}(\omega) = {\cal U}_{{\rm m}}(0, \omega)$, ${\cal U}_{{\rm m}2}(\omega) = {\cal U}_{{\rm m}}(-d_{\rm F}, \omega)$ refer to values taken at the interfaces with the normal metals N1 and N2. From Eqs.~\eqref{eq:ferromagnetlinresponse} and \eqref{eq:ferromagnetdiffsolution}, we relate spin and heat currents with the magnon chemical potential and the magnon temperature,
\begin{align}
\begin{split}
    (-1)^{\ia-1} \bigg[ \cosh{(\Lambda(\omega)^{-1} d_{\rm F})}
    {\cal U}_{{\rm m}\ia }(\omega)
    - 
    {\cal U}_{{\rm m}\ib }(\omega) \bigg] & \\ =
    - \sinh(\Lambda(\omega)^{-1} d_{\rm F})
    \mathcal{Z}_{{\rm F}\parallel}^{\infty} (\omega)
    {\cal I}_\ia(\omega) &
\end{split}
\label{eq:ferromagnonheatrelation}
\end{align}
for $\ia \neq \ib$, where 
\begin{equation}
    \mathcal{Z}_{{\rm F}\parallel}^{\infty} (\omega)
  = \Lambda(\omega) \Sigma(\omega)^{-1} .
\label{eq:zfparainf}
\end{equation}

\section{Magnon-mediated current drag}
\label{sec:conductivities}
In Sec.\ \ref{sec:linearresponse} we established linear relations between (generalized) currents and voltages for each part of the N$|$F$|$N trilayer separately. When combining these relations, they constitute a system of equations, from which we can deduce the linear response of the trilayer to the applied electric fields $E_1(\omega)$ and $E_2(\omega)$ in N1 and N2. In this Section, we  discuss the charge current response, which is characterized by the linear conductivities $\sigma^{xx}_{\ib\ia}(\omega)$ and $\sigma^{yx}_{\ib\ia}(\omega)$ of Eqs.~\eqref{eq:linear_conductivity_definition_x} and \eqref{eq:linear_conductivity_definition_y}, respectively. The thermal response will be discussed in Sec.\ \ref{sec:peltier}.

\subsection{Trilayer response}

Solving the coupled Equations \eqref{eq:nmspinrelation}, \eqref{eq:fnspinrelationperp}, and \eqref{eq:fispinrelationperp} for the transverse components of the spin accumulations $\vu_{{\rm s}\ia }$ at each interface $\ia= 1,2$, we find a linear relationship between the transverse spin accumulation $u_{{\rm s}\ia \perp}(\omega)$ at the interface and the source terms $\delta \vu_{{\rm s}\ib }(\omega)$ of Eq.\ (\ref{eq:linearsourceterm}),
\begin{align}
\label{eq:linresponse_perp}
\begin{split}
  u_{{\rm s}\ia  \perp}(\omega) =&\, \sum_{\ib=1}^{2} \zzeta_{\ia\ib\perp}(\omega)
  \delta u_{{\rm s}\ib  \perp}(\omega).
\end{split}
\end{align}
The dimensionless response coefficients $f_{\ia\ib\perp}(\omega)$ depend on the impedances $Z_{{\rm N}\ia }$, $Z_{{\rm FN}\ia \perp}$, $Z_{{\rm F}\perp}^{\infty}(\omega)$, and on the product $k(\omega) d_{\rm F}$.
For the explicit expression we refer to App.~\ref{sec:impedances}.
Similarly, for the longitudinal component of the spin current and the heat current at the interfaces we find from Eqs.~(\ref{eq:nmspinrelation}), (\ref{eq:nmheatrelation}), \eqref{eq:fnrelationpara}, and \eqref{eq:ferromagnonheatrelation} that
\begin{equation}
    {\cal U}_{{\rm e}\ia }(\omega) = \sum_{\ib=1}^{2} \calzzeta_{\ia\ib}(\omega) \delta {\cal U}_{{\rm e}\ib }(\omega),
\label{eq:linresponse_para}
\end{equation}
where the dimensionless $2 \times 2$ matrices $\calzzeta_{\ia\ib}(\omega)$ depend on the impedances ${\cal Z}_{{\rm N}\ia }$, ${\cal Z}_{{\rm FN}\ia \parallel}$, ${\cal Z}_{{\rm F}\parallel}^{\infty}(\omega)$, and on the product $\Lambda(\omega)^{-1} d_{\rm F}$, see App.\ \ref{sec:impedances}.

Using the source term \eqref{eq:linearsourceterm}, we may then calculate the spin accumulation in response to an applied field,
\begin{align}
    \vu_{{\rm s}\ia }(\omega) =&\ 2 \sum_{\ib=1}^{2} (-1)^{\ib-1} \lambda_{{\rm N}\ib }(\omega) \theta_{{\rm SH}\ib } E_\ib 
  \nonumber \\ &\, \mbox{} \times
  \left[ \calzzeta_{\ia\ib}(\omega)_{11} (\ve_y \cdot \vm_{\rm eq}) \vm_{\rm eq}
  \right. \nonumber \\ &\, \ \ \ \ \left. \mbox{}
 + \zzeta_{\ia\ib\perp}(\omega) (\ve_y \cdot \ve_{\perp}^*) \ve_{\perp}
  \right. \nonumber \\ &\, \left. \ \ \ \ \mbox{}
 + \zzeta_{\ia\ib\perp}(-\omega)^* (\ve_y \cdot \ve_{\perp}) \ve_{\perp}^* \right],
\end{align}
where the index $11$ of $\calzzeta_{\ia\ib}(\omega)_{11}$ points to the matrix element of the $2 \times 2$ matrix. From here, we find the local and nonlocal conductivities 
\begin{align}
  \label{eq:sigmaxx}
  \begin{split}
  \frac{\sigma^{xx}_{\ib\ia}(\omega)}{\sigma_{{\rm N}\ib }} =&\ 
  \delta_{\ib\ia}\left[1 + \frac{\lambda_{{\rm N}\ia }}{d_{{\rm N}\ib }} \theta_{{\rm SH}\ia } \theta_{{\rm SH}\ib } \right] \\
  &\, + \frac{\lambda_{{\rm N}\ia }}{2 d_{{\rm N}\ib }}
  \left[ s_{\ib\ia}(\omega) (1 - m_y^2) + s_{\ib\ia}'(\omega)\right], 
  \end{split} \\
  \label{eq:sigmayx}
  \frac{\sigma^{yx}_{\ib\ia}(\omega)}{\sigma_{{\rm N}\ib }} =&\ 
  \frac{\lambda_{{\rm N}\ia }}{2 d_{{\rm N}\ib }}
  \left[ s_{\ib\ia}(\omega) m_x m_y + s_{\ib\ia}''(\omega) m_z \right],
\end{align}
where we used Eqs.~\eqref{eq:equilibriummagnetization} and \eqref{eq:eperp} for the unit vectors $\vm_{\rm eq}$ and $\ve_{\perp}$ and we introduced the dimensionless response coefficients
\begin{align}
  \label{eq:sji}
  s_{\ib\ia}(\omega) =&\ 
  (-1)^{\ia+\ib} \theta_{{\rm SH}\ia } \theta_{{\rm SH}\ib }
  [\zzeta_{\ib\ia\perp}(\omega) + \zzeta_{\ib\ia\perp}(-\omega)^* \nonumber \\
  &\ \quad \quad  \quad \quad  \quad \quad  \quad \ - 2 \calzzeta_{\ib\ia}(\omega)_{11}], \nonumber \\
  s_{\ib\ia}'(\omega) =&\ (-1)^{\ia+\ib} 2 \theta_{{\rm SH}\ia } \theta_{{\rm SH}\ib }
  [\calzzeta_{\ib\ia}(\omega)_{11} - \delta_{\ib\ia}] ,
  \\ \nonumber
  s_{\ib\ia}''(\omega) =&\ (-1)^{\ia+\ib} i \theta_{{\rm SH}\ia } \theta_{{\rm SH}\ib }
  [\zzeta_{\ib\ia\perp}(\omega) - \zzeta_{\ib\ia\perp}(-\omega)^*],
\end{align}
which each describe contributions to the conductivity with different characteristic dependences on the magnetization direction $\vm_{\rm eq}$ \cite{Cornelissen2015-fh, Goennenwein2015-lb, Li2016-ye, Wu2016-cs}.

Equations (\ref{eq:sigmaxx})--(\ref{eq:sji}) are a central result of this article. They generalize the theory of Ref.\ \onlinecite{Reiss2021-em} to the nonlocal response and go beyond Ref.\ \onlinecite{Reiss2021-em} by including magnon relaxation processes in F and the coupling of spin and heat transport across the F$|$N interfaces.
The term proportional to $m_x m_y$ in Eq.\ (\ref{eq:sigmayx}), with the dimensionless off-diagonal conductivity coefficient $s^{xy}_{11}$, can be identified with a spin-Hall version of the planar Hall effect, which is symmetric under magnetization reversal. The terms proportional to $m_z$ correspond to a spin-Hall version of the anomalous Hall effect, which is antisymmetric under magnetization reversal. In the low-frequency limit, Eqs.\ (\ref{eq:sigmaxx})--(\ref{eq:sji}) reproduce existing results by Refs.~\onlinecite{Cornelissen2016-wy,Zhang2012-fy}. Upon sending the interface transparency between F and N2 to zero, the trilayer considered here effectively becomes a bilayer, and our results reproduce previous results of the bilayer spin-Hall magnetoresistance of Refs.\ \onlinecite{Chen2013-gf,Chen2016-pc,Cornelissen2016-wy,Zhang2019-zv,Zhang2012-fy,Zhang2012-ig,Wang2018-qd} in the zero-frequency limit and Ref.\ \onlinecite{Reiss2021-em} for nonzero frequency.

The frequency dependence of the local and nonlocal spin-Hall conductivity arises from the frequency dependence of the response of the F layer. The response of the F$|$N interfaces and the normal metals N$\ia$ is frequency independent for frequencies well into the THz range. The frequency dependence of the transverse contribution, represented by the coefficient $\zzeta_{\ib\ia\perp}$, reflects the excitation of coherent magnetization modes, starting with the uniform ferromagnetic resonance mode \cite{Jiao2013-im, Wei2014-uu}, and including non-uniform magnon modes for higher frequencies \cite{Reiss2021-em}. The frequency dependence of the longitudinal contribution arises from the diffusive motion of incoherent, thermal magnons through the F layer and typically already sets in at lower frequencies. 

\subsection{Numerical estimates}
\label{sec:numericalestimates}

\begin{figure}
     \centering
     \includegraphics[width=0.5\textwidth]{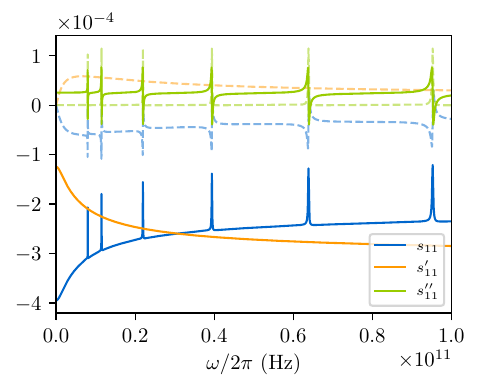}
     \caption{Real part (solid line) and imaginary part (dashed line) of local linear response coefficients $s_{11}(\omega)$ (blue), $s_{11}'(\omega)$ (orange), and $s_{11}''(\omega)$ (green), see Eq.~\eqref{eq:sji}. The three dimensionless coefficients have a different dependence on the direction of the magnetization in F and characterize the local conductivity correction from the coupling of N to F$|$N. We note that the coefficient $s_{11}(\omega)$ is directly proportional to the SMR response calculated in Ref.~\onlinecite{Reiss2021-em}, while $s_{11}''(\omega)$ describes the anomalous Hall effect (AHE).}
     \label{fig:s11}
\end{figure}

\begin{figure}
     \centering
     \includegraphics[width=0.5\textwidth]{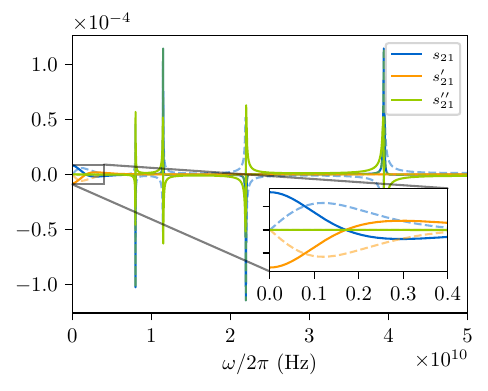}
     \caption{Magnon-mediated electric current drag. Real part (solid line) and imaginary part (dashed line) of nonlocal linear response coefficients $s_{21}(\omega)$ (blue), $s_{21}'(\omega)$ (orange), and $s_{21}''(\omega)$ (green), see Eq.~\eqref{eq:sji}. At low frequency, the nonlocal conductivity is mediated by thermal magnons, but coherent spin waves dominate at GHz to THz frequencies.}
     \label{fig:s21}
\end{figure}

\begin{figure}
     \centering
     \includegraphics[width=0.5\textwidth]{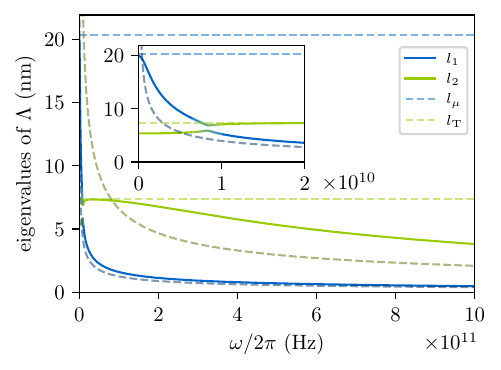}
     \caption{Eigenvalues $l_1$ and $l_2$ of $\Lambda$, Eq.~\eqref{eq:capitallambda}. The eigenvalues at zero frequency correspond to the characteristic decay lengths $l_{\mu}$, Eq.~\eqref{eq:approxlmu}, and $l_{\rm T}$, Eq.~\eqref{eq:approxlT}, for magnon or heat transport, respectively. Their numerical values are given in Tab.~\ref{tab:derivedparameters}. We refer to Sec.~\ref{sec:numericalestimates} for an estimate of $l_{\rm m} (\omega)$, $m=1,2$, in the large-frequency limit.}
     \label{fig:LambdaEig}
\end{figure}

We now present characteristic numerical estimates for the local and nonlocal spin-Hall magnetoresistance effect, using material and device parameters for a Pt$|$YIG$|$Pt-trilayer. (We regard the ferrimagnet yttrium iron garnet (YIG) as ferromagnetic.) The material and device parameters used are summarized in Tabs.\ \ref{tab:materialparameters} and \ref{tab:derivedparameters}. We assume equal thickness $d_{{\rm N}1} = d_{{\rm N}2}$ for the two Pt layers.
The dimensionless coefficients $s_{\ib\ia}(\omega)$, $s'_{\ib\ia}(\omega)$, and $s''_{\ib\ia}(\omega)$, which each describe a contribution to the local and nonlocal conductivities with a different characteristic magnetization dependence, are shown in Figs.~\ref{fig:s11} and \ref{fig:s21} as a function of the driving frequency $\omega$.
The frequency dependence of the longitudinal response is dominated by the two lengths $l_{m}(\omega)$, $m=1,2$, for magnon spin and energy relaxation, which are the two eigenvalues of the $2 \times 2$ matrix $\Lambda(\omega)$, see Eq.~(\ref{eq:capitallambda}). The frequency dependence of these two relaxation lengths is shown separately in Fig.~\ref{fig:LambdaEig}.

One of the key features of the spin-Hall magnetoresistance effect is that the electrical conductivity depends on the direction of the magnetization in F, see Eqs.~\eqref{eq:sigmaxx} and \eqref{eq:sigmayx}. Specifically, both $\sigma^{xx}_{\ib\ia}$ and $\sigma^{yx}_{\ib\ia}$ show a characteristic sine-like behavior with respect to the angle $\theta$ between the magnetization direction and the $y$-axis. While $\sigma^{yx}_{\ib\ia}$ changes sign as a function of the magnetization direction, the correction to the local conductivities $\sigma^{xx}_{11}$ and $\sigma^{xx}_{22}$ are strictly positive for zero frequency. In the zero-frequency limit, the nonlocal conductivities $\sigma^{xx}_{12}$ and $\sigma^{xx}_{21}$ are negative, except for a magnetization in the $xz$ plane, for which $\sigma^{xx}_{12} = \sigma^{xx}_{21}$ vanishes. Hence, the nonlocal response of the N$|$F$|$N trilayer is such that it can act as a pure spin valve, being able to switch the nonlocal response on and off, if it is possible to rotate the magnetization into and out of the $xz$ plane, {\em e.g.}, via an external magnetic field.

{\renewcommand{\arraystretch}{1.3}
\begin{table}
\begin{ruledtabular}
    \centering
    \begin{tabular}{llr}
    \textrm{Quantity} & \textrm{Value} & \textrm{Ref.} \\
    \colrule
    $T$ & $ 300 \, \mathrm{K}$ & - \\
    $g_{\uparrow\downarrow} $ & $(6 + 0.3i) \times 10^{13} \, \Omega^{-1} \, \mathrm{m}^{-2}$ & \onlinecite{Qiu2013-rv, Althammer2013-zm, Hahn2013-rw} \\
    \hline
    \multicolumn{3}{c}{YIG} \\
    \hline
    $d_{\rm F}$ & $6 \times 10^{-8} \, \mathrm{m}$ & - \\
    $\alpha$ & $2 \times 10^{-4}$ & \onlinecite{Cornelissen2015-fh} \\
    $\omega_0 / 2 \pi$ & $ 8 \times 10^{9} \, \mathrm{s}^{-1}$ & \onlinecite{Hahn2013-rw} \\
    $D_{\rm ex}$ & $8 \times 10^{-6} \, \mathrm{m}^2 \, s^{-1}$ & \onlinecite{Weiler2013-ny} \\
    $a$ & $ 1.2376 \times 10^{-9} \, \mathrm{m}$ & \onlinecite{Cherepanov1993-sr} \\
    $s a^3$ & $10$ & \onlinecite{Cherepanov1993-sr} \\
    $\tau_{\rm m,el}$ & $ 0.11 \, \mathrm{ps}$ & - \\
    $\tau_{\rm m,ex}$ & $ 2.7 \, \mathrm{ps}$ & \onlinecite{Shi2021-kc} \\
    $\tau_{\rm mp,ex}$ & $ 0.5 \, \mathrm{ps}$ & \onlinecite{Cornelissen2016-wy} \\
    $\tau_{\rm m,rel}, \ \tau_{\rm mp,rel}$ & $ 2 \hbar / (\alpha k_{\rm B} T) \approx 255 \, \mathrm{ps}$ & - \\
    \hline
    \multicolumn{3}{c}{Pt} \\
    \hline
    $d_{\rm N}$ & $ 4 \times 10^{-9} \, \mathrm{m}$ & - \\
    $\theta_{\rm SH}$ & $ 0.1 $ & \onlinecite{Althammer2013-zm} \\
    $\lambda_{\rm N}$ & $ 2 \times 10^{-9} \, \mathrm{m}$ & \onlinecite{Weiler2013-ny} \\
    $\sigma_{\rm N}$ & $ 9 \times 10^{6} \, \Omega^{-1} \, \mathrm{m}^{-1}$ & \onlinecite{Corti1984-xd} \\
    $C_{\rm p}$ & $ 2.73 \times 10^{6} \, \mathrm{J} \, \mathrm{K}^{-1} \, \mathrm{m}^{-3}$ & \onlinecite{Lin2008-tz, Rumble2022-ev} \\
    $C_{\rm e}$ & $ 0.13 \times 10^{6} \, \mathrm{J} \, \mathrm{K}^{-1} \, \mathrm{m}^{-3}$ & \onlinecite{Lin2008-tz, Rumble2022-ev, Sullivan2023-mw} \\
    $l_{\rm ep}$ & $ 4.5 \times 10^{-9} \, \mathrm{m}$ & \onlinecite{Flipse2014-kh, Weiler2013-ny} \\
    \end{tabular}
    \caption{Material parameters for YIG and Pt. The electron-phonon time $\tau_{\rm ep} \approx 40 \, \mathrm{fs}$ in Eq.~\eqref{eq:heatcontinuity} is calculated from $l_{\rm ep}$ and $\kappa_{\rm e}$, which is in turn calculated from $\sigma_{\rm N}$ via the Wiedemann-Franz law. The time scale $\tau_{\rm m,ex}$ is mainly associated with ``four-magnon scattering'' and we estimate the combined relaxation times $\tau_{\rm m,rel}$ and $\tau_{\rm mp,rel}$ from the Landau-Lifshitz-Gilbert equation, since both processes are not magnon-number conserving. At room temperature, spin-non-conserving ``three-magnon scattering'' can be neglected in comparison to the spin-conserving ``four-magnon scattering'' \cite{Shi2021-kc}. The impurity scattering $\tau_{\rm m,el}$ is fitted to reproduce the magnon conductivity \mbox{$\sigma_{\rm m} \approx 4 \times 10^5 \, \mathrm{S} \mathrm{m}^{-1}$} measured in Ref.~\onlinecite{Cornelissen2015-fh}. A comparison to a microscopical model as discussed in Ref.~\onlinecite{Schmidt2021-qo} yields a similar order of magnitude for the relaxation time at room temperature.}
    \label{tab:materialparameters}
\end{ruledtabular}
\end{table}}

{\renewcommand{\arraystretch}{1.3}
\begin{table}
\begin{ruledtabular}
    \centering
    \begin{tabular}{lllr}
    \multicolumn{2}{l}{\textrm{Quantity}} & \textrm{Value} & \textrm{Eq.} \\
    \colrule
    $l_{\mu}$ & & $20 \, \mathrm{nm}$ & \eqref{eq:approxlmu} \\
    $l_{\rm T}$ & & $7 \, \mathrm{nm}$ & \eqref{eq:approxlT} \\
    \hline
    \multirow{4}{*}{$\mathcal{C}_{\rm m}$}
    & $[{\cal C}_{\rm m}]_{11}$ & $150 \, \mathrm{ns} \, \Omega^{-1} \, \mu\mathrm{m}^{-3}$ & \eqref{eq:calCm} \\
    & $[{\cal C}_{\rm m}]_{12}$ & $11 \, \mathrm{ns} \, \Omega^{-1} \, \mu\mathrm{m}^{-3}$ & \eqref{eq:calCm} \\
    & $[{\cal C}_{\rm m}]_{21}$ & $11 \, \mathrm{ns} \, \Omega^{-1} \, \mu\mathrm{m}^{-3}$ & \eqref{eq:calCm} \\
    & $[{\cal C}_{\rm m}]_{22}$ & $15 \, \mathrm{ns} \, \Omega^{-1} \, \mu\mathrm{m}^{-3}$ & \eqref{eq:calCm} \\
    \hline
    \multicolumn{2}{l}{$c = (k_{\rm B}^2 T / 2 e^2) [{\cal C}_{\rm m}]_{22}  / \rho_{\rm YIG}$} & $3.3 \, \mathrm{J} \, \mathrm{kg}^{-1} \, \mathrm{K}^{-1}$ & \eqref{eq:calCm} \\
    \hline
    \multirow{4}{*}{$\mathcal{G}_{\rm m}$}
    & $[{\cal G}_{\rm m}]_{11}$ & $11 \times 10^{2} \, \Omega^{-1} \, \mu\mathrm{m}^{-3}$ & \eqref{eq:calGmcalCmrelation} \\
    & $[{\cal G}_{\rm m}]_{12}$ & $45 \times 10^{0} \, \Omega^{-1} \, \mu\mathrm{m}^{-3}$ & \eqref{eq:calGmcalCmrelation} \\
    & $[{\cal G}_{\rm m}]_{21}$ & $45 \times 10^{0 } \, \Omega^{-1} \, \mu\mathrm{m}^{-3}$ & \eqref{eq:calGmcalCmrelation} \\
    & $[{\cal G}_{\rm m}]_{22}$ & $29 \times 10^{3} \, \Omega^{-1} \, \mu\mathrm{m}^{-3}$ & \eqref{eq:calGmcalCmrelation} \\
    \hline
    \multirow{4}{*}{$\Sigma_{\rm m}(0)$}
    & $[\Sigma_{\rm m}]_{11} = \sigma_{\rm m}$ & $0.41 \, \Omega^{-1} \, \mu \mathrm{m}^{-1}$ & \eqref{eq:sigmam} \\
    & $[\Sigma_{\rm m}]_{12} = e L_{\rm m} / k_{\rm B} T$ & $0.55 \, \Omega^{-1} \, \mu \mathrm{m}^{-1}$ & \eqref{eq:L} \\
    & $[\Sigma_{\rm m}]_{21} = e L_{\rm m} / k_{\rm B} T$ & $0.55 \, \Omega^{-1} \, \mu \mathrm{m}^{-1}$ & \eqref{eq:L} \\
    & $[\Sigma_{\rm m}]_{22} = 2 e^2 \kappa_{\rm m} /  k_{\rm B}^2 T$ & $1.63 \, \Omega^{-1} \, \mu \mathrm{m}^{-1}$ & \eqref{eq:kappam} \\
    \hline
    $\sigma_{\rm m}$ & & $4.1 \times 10^5 \, \mathrm{S} \, \mathrm{m}^{-1}$ & \eqref{eq:sigmam} \\
    $L_{\rm m}$ & & $1.4 \times 10^4 \, \mathrm{A} \, \mathrm{m}^{-1}$ & \eqref{eq:L} \\
    $\kappa_{\rm m}$ & & $1.8 \, \mathrm{W} \, \mathrm{K}^{-1} \, \mathrm{m}^{-1}$ & \eqref{eq:kappam} \\
    \end{tabular}
    \caption{Derived quantities for YIG from the Boltzmann theory in App.~\ref{sec:boltzmann}. The characteristic length scales associated with magnon transport mediated (mainly) by the magnon chemical potential or magnon temperature are the eigenvalues of $\Lambda$, Eq.~\eqref{eq:capitallambda}, at zero frequency, $l_1 \sim l_{\mu}$ and $l_2 \sim l_{\rm T}$, respectively (see Sec.~\ref{sec:numericalestimates}). Consistent with Ref.~\onlinecite{Cornelissen2016-wy}, our model suggests that the magnon chemical potential carries farther than the magnon temperature. In general, Refs.~\onlinecite{Cornelissen2016-wy, Cornelissen2015-fh, Sterk2019-mt} suggest a longer magnon spin diffusion length than the value listed here. We attribute this difference to a shortcoming of the relaxation time approximation in the Boltzmann theory that neglects subthermal magnon transport, as discussed in Ref.~\onlinecite{Boona2014-cm}.}
    \label{tab:derivedparameters}
\end{ruledtabular}
\end{table}}

\subsection{Analytical estimates}

We now give order-of-magnitude estimates for the magnon relaxation lengths $l_{m}(\omega)$, $m=1,2$, which are the eigenvalues of the $2 \times 2$ matrix $\Lambda(\omega)$. 
Order-of-magnitudes for the spin impedances are given in App.\ \ref{sec:analytical}. 

{\it In the large-frequency limit}, we may approximate
\begin{align}
  \Lambda(\omega) \approx \frac{e^{i \pi/4}}{\sqrt{\omega}}
         {\cal C}_{\rm m}^{-1/2} \Sigma_{\rm m}^{1/2},
\label{eq:approxcapitallambda}
\end{align}
where $\omega > 0$ and $\Lambda(\omega) = \Lambda(-\omega)^*$. Here, we can neglect the frequency-dependence of $\Sigma_{\rm m}(\omega)$ for frequencies smaller than the magnon momentum-relaxation, $\omega < \tau_{\rm m}^{-1}$, which has a time scale of $\tau_{\rm m} < 0.1 \, \mathrm{ps}$. Since the four entries of the response matrix $\Sigma_{\rm m}$ are all of comparable magnitude --- all four entries are governed by the momentum relaxation time $\tau_{\rm m}$ ---, the two relaxation lengths,
\begin{align}
    l_{1}(\omega) &\sim (\Dex \tau_{\rm m} k_{\rm B} T/\hbar \omega)^{1/2}(\hbar \omega_0/k_{\rm B} T)^{1/4}, \\
    l_{2}(\omega) &\sim (\Dex \tau_{\rm m} k_{\rm B} T/\hbar \omega)^{1/2},
\end{align}
are determined by the two principal values of the response matrix ${\cal C}_{\rm m}$, which differ by a factor $\sim \sqrt{k_{\rm B} T/\hbar \omega_0}$, see App. \ref{sec:boltzmann}.
As seen in Fig.~\ref{fig:LambdaEig}, however, for the longer relaxation length the asymptotic frequency dependence sets in only for frequencies well above $1 \, \mathrm{THz}$.

The transverse response is dominated by the impedances of the two ferromagnet--normal-metal interfaces, except in the vicinity of resonance frequencies
\begin{equation}
  \omega_n = \Dex \left( \frac{n \pi}{d_{\rm F}} \right)^2 + \omega_0,
  \label{eq:resonancefrequencies}
\end{equation}
where it shows a sharply peaked behavior as a function of the driving frequency $\omega$.

{\it For the low-frequency limit} of the longitudinal response, the relaxation lengths $l_{m}(\omega)$ for magnon spin and energy density depend not only on the momentum relaxation time $\tau_{\rm m}$, but also on the much longer relaxation times $\tau_{{\rm mp},{\rm ex}}$ (exchange-based magnon-phonon scattering), $\tau_{{\rm mp},{\rm rel}}$ (spin-nonconserving magnon-phonon scattering), and $\tau_{{\rm m},{\rm rel}}$ (spin-nonconserving magnon-magnon scattering). Of these, relaxation by exchange-based magnon-phonon scattering is dominant at room temperature. Since exchange-based magnon-phonon scattering conserves the magnon spin density, but not the magnon energy density, the smallest relaxation length $l_2(0) \sim l_{\rm T}$ describes the relaxation of the magnon temperature, whereas the relaxation length $l_1(0) \sim l_{\mu}$ for the magnon chemical potential is much larger. To find an order-of-magnitude estimate, we neglect the off-diagonal elements of the matrices ${\cal G}(\omega)$ and $\Sigma(\omega)$ in Eq.\ \eqref{eq:capitallambda} and find (compare with Ref.~\onlinecite{Cornelissen2016-wy}),
\begin{align}
    l_{\mu} &\equiv ([\mathcal{G}_{\rm m}^{-1}]_{11} [\Sigma]_{11})^{1/2} \label{eq:approxlmu}  \\
    &\approx \left(2 \Dex \tau_{\rm m} \tau_{\rm rel} k_{\rm B} T \zeta (3/2)/\sqrt{\pi} \hbar \right)^{1/2} (\hbar \omega_0 / k_{\rm B} T)^{1/4}, 
    \nonumber
    \\
    \begin{split}
    l_{\rm T} &\equiv ([\mathcal{G}_{\rm m}^{-1}]_{22} [\Sigma]_{22})^{1/2} \\
    &\approx \left( 4 \Dex \tau_{\rm m} \tau_{{\rm mp},{\rm ex}} k_{\rm B} T/\hbar\right)^{1/2}.
    \label{eq:approxlT}
    \end{split}
\end{align}
Here, $\tau_{\rm rel} = \min(\tau_{{\rm m},{\rm rel}},\tau_{{\rm mp},{\rm rel}})$ and the subscripts ``11'' and ``22'' point to the upper left and lower right matrix elements of the $2 \times 2$ matrices, respectively. The values of $l_{\mu}$ and $l_{\rm T}$ are shown in Tab.~\ref{tab:derivedparameters} and are close to the actual eigenvalues of $\Lambda$, see Fig.~\ref{fig:LambdaEig}.

\section{Spin-Peltier effect}
\label{sec:peltier}
\begin{figure}
  \centering
  \includegraphics[width=0.38\textwidth]{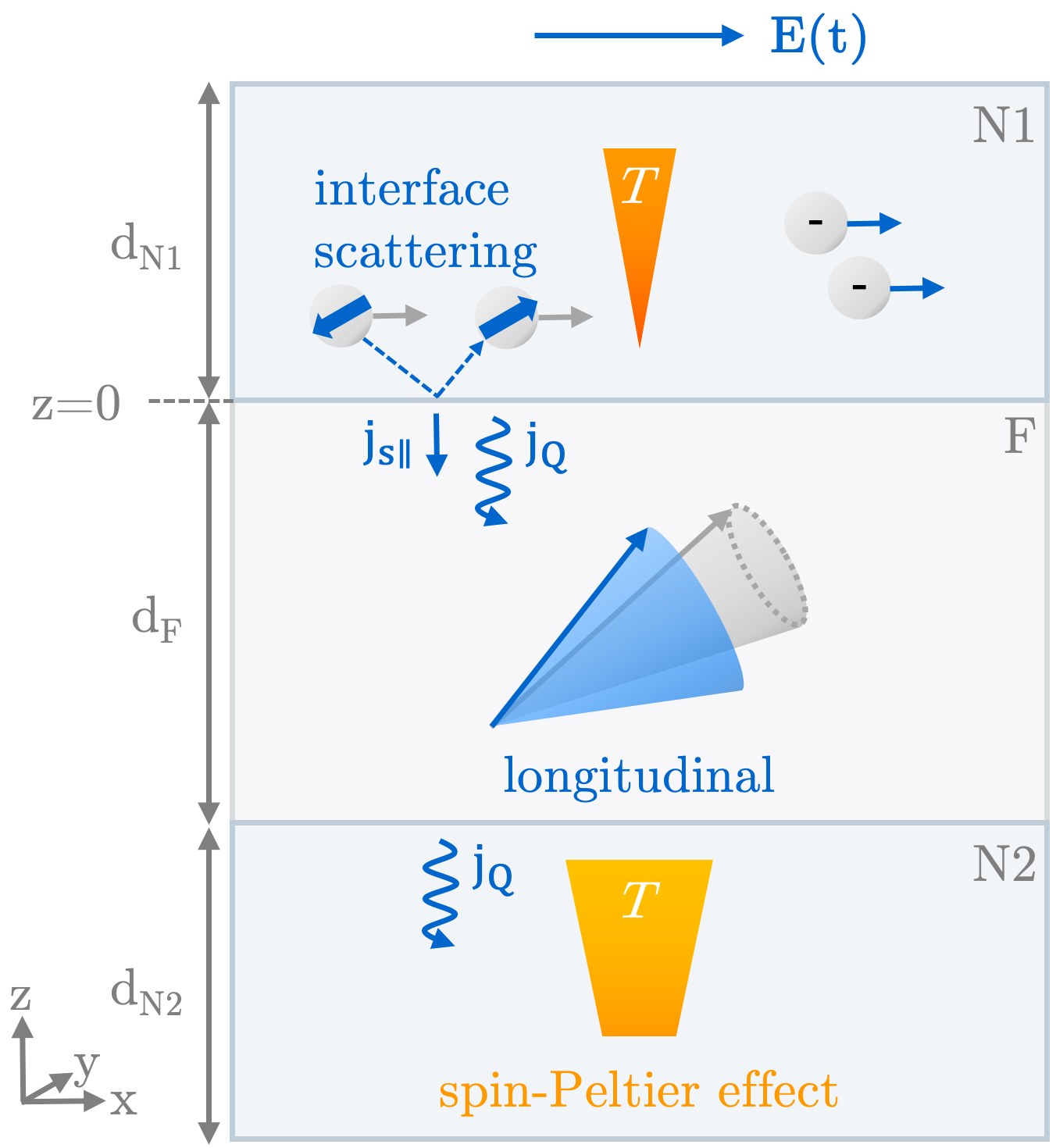}
\caption{Local and nonlocal spin Peltier effect in an N$|$F$|$N trilayer. Since longitudinal spin and heat currents are coupled inside F and at the two N$|$F interfaces, the flow of a spin current linear in the applied electric field also implies the flow of a heat current and, hence, the generation of a temperature gradient.}
\label{fig:peltier}
\end{figure}

A consequence of the linear coupling of spin and heat currents across the F$|$N interfaces is that heat currents can be driven into or out of the normal metal to linear order in the applied electric field, see Fig.\ \ref{fig:peltier}. The sign of these heat currents depends on the sign of the electric field or the magnetization in F, so that not only current-induced heating, but also current-induced cooling of electrons in N1 or N2 is possible (at the expense of heating up the magnons in F and/or the other normal-metal layer) \cite{Flipse2014-kh}. This makes it possible to construct thermoelectric heaters or coolers based on the coupling between spin and heat transport in ferromagnetic-insulator/normal-metal heterostructures. 

The linear response relation (\ref{eq:linresponse_para}), combined with Eq.\ (\ref{eq:linearsourceterm}) for the source term $\delta u_{{\rm s}\ia }$, gives the change $\Delta T_{{\rm e}\ia } = (e/k_{\rm B}) u_{{\rm eQ}\ia }$ of the electron temperature at the F$|$N$\ia$ interface to linear order in the applied electric field,
\begin{align}
  \Delta T_{{\rm e}\ia } (\omega) = &\, \frac{2 e}{k_{\rm B}} \sum_{\ib=1}^2 (-1)^{\ib-1} \calzzeta_{\ia\ib} (\omega)_{21}
    \lambda_{{\rm N}\ib } \theta_{{\rm SH}\ib } E_\ib(\omega) m_y.
    \label{eq:dTPeltier}
\end{align}
The spatially averaged temperature change $\overline{\Delta T_{{\rm e}\ia }}(\omega)$ in N$\ia$, $\ia=1,2$, is then given by Eq.\ (\ref{eq:Peltiereffect}), with
\begin{align}
  \eta_{\ib\ia}^x(\omega) =&\,
  \frac{2 e}{k_{\rm B}} (-1)^{\ia-1} \calzzeta_{\ib\ia} (\omega)_{21}
    \lambda_{{\rm N}\ia } \theta_{{\rm SH}\ia } m_y \nonumber \\ &\, \mbox{} \times
    \frac{l_{{\rm ep},\ib }(\omega)}{d_{{\rm N}\ib }}
    \tanh \frac{d_{{\rm N}\ib }}{l_{{\rm ep},\ib }(\omega)}
    .
\end{align}

The maximum cooling that can be obtained in such a spin-Peltier effect is bounded by competition with Joule heating from charge currents in N. This is an effect quadratic in the applied field, which is considered in detail in our companion article Ref.\ \onlinecite{Franke2025-nonlin}. An estimate in the zero-frequency limit $E_\ia(t) = E_\ia$ can already be obtained from the linear theory developed here. Neglecting the heating from dissipative spin currents in the N layers \cite{Tulapurkar2011-lm, Taniguchi2016-ei}, which is quadratic in the spin-Hall angle, the Joule heating rate $\sigma_{{\rm N}\ia } E_\ia^2$ implies a temperature change $\delta T_{{\rm e}\ia } = (e/k_{\rm B}) \delta u_{{\rm eQ}\ia }$ at the F$|$N$\ia$ interface, where
\begin{equation}
  \delta u_{{\rm eQ}\ia } = \frac{k_{\rm B}}{e}
  \frac{\sigma_{{\rm N}\ia } \tau_{{\rm ep},\ia }}{C_{{\rm e}\ia }} E^2_\ia,
\end{equation}
with $C_{{\rm e}\ia }$ the electronic heat capacity of N$\ia$, $\ia=1,2$.
The interfacial temperature change from Joule heating can be used as a source term in the linear-response relation (\ref{eq:linresponse_para}), which gives the quadratic-in-$E$ contribution to the change of the electronic temperature at the F$|$N$\ib$ interface,
\begin{equation}
  \Delta T_{{\rm e}\ib }^{(2)} =
  \sum_{\ia=1}^{2} {\cal F}_{\ib\ia}(0)_{22}
  \frac{\sigma_{{\rm N}\ia } \tau_{{\rm ep},\ia }}{C_{{\rm e}\ia }} E_\ia^2,
  \label{eq:dTJoule}
\end{equation}
where the linear-response coefficient ${\cal F}_{\ib\ia}(\omega)$ was defined in Eq.\ (\ref{eq:linresponse_para}).

\begin{figure}
    \centering
    \includegraphics[width=0.5\textwidth]{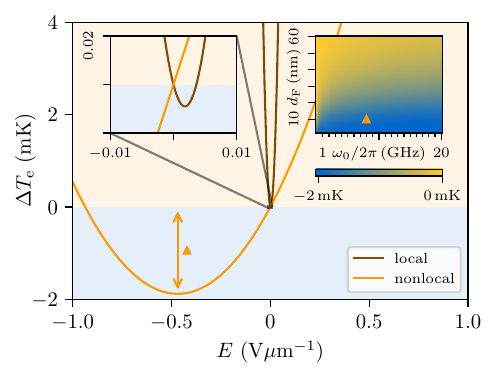}
    \caption{Temperature change $\Delta T_{{\rm e}\ia }$ at the F$|$N$\ia$ interface vs.\ applied electric field $E_1$ for $\vm_{\rm eq} = \ve_y$ and for material and device parameters listed in Tabs.\ \ref{tab:materialparameters} and \ref{tab:derivedparameters}, except for the thickness $d_{\rm F} = 10 \, \mathrm{nm}$. The maximum cooling temperature $\blacktriangle$ is determined by the competition of linear-in-$E$ spin-Peltier cooling and quadratic-in-$E$ Joule heating. The insets show a zoom-in for $\Delta T_{\rm e}$ near $E = 0$ and the maximum nonlocal cooling as a function of magnetic field $\omega_0$ and thickness $d_{\rm F}$.}
\label{fig:peltier2}
\end{figure}

Figure \ref{fig:peltier2} shows the temperature change $\Delta T_{{\rm e}\ia }$ at the two F$|$N interfaces for a Pt$|$YIG$|$Pt trilayer as a function of the applied electric field $E_1 = E$ in N1, taking into account the linear-in-$E$ and quadratic-in-$E$ temperature changes of Eqs.\ (\ref{eq:dTPeltier}) and (\ref{eq:dTJoule}). Material and device parameters are taken from Tabs.\ \ref{tab:materialparameters} and \ref{tab:derivedparameters}. Our semi-phenomenological transport equations include heat exchange with the phonon bath, assuming a phonon heat capacity large enough that the phonon bath temperature stays constant, see Sec.~\ref{sec:linearresponse}. The estimates of Fig.\ \ref{fig:peltier2} do not consider phonon-mediated heat transport through the F layer, which is an additional nonlocal heating source and, hence, limits the maximal temperature change than can be achieved in the nonlocal setup. 

Although the nonlocal linear-in-$E$ cooling effect via the spin-Peltier effect is smaller than the local effect, the difference between the temperature changes from nonlocal and local Joule heating is even larger --- both nonlocal effects are magnon-mediated ---, so that the maximum cooling effect in the nonlocal setup (electric field and temperature change are in different N layers) is larger than in the local setup (electric field and temperature change in the same layer). The smallness of the nonlocal Joule heating effect comes in part, because the relaxation length for magnon heat transport in F is smaller than that for magnon spin transport, see Fig.\ \ref{fig:LambdaEig}. However, both heat and spin currents have the same asymptotic dependence on $d_{\rm F}$ because of the coupling of heat and spin currents across the F$|$N interfaces. 

The local spin-Peltier effect has been observed by \citet{Flipse2014-kh} for Pt$|$YIG bilayers.
Thermal imaging by \citet{Daimon2016-kt} underscores that the spin-Peltier effect allows pinpoint modulation of the temperature in the mK regime, consistent with the estimates given above. The spatial separation of spin injector and heat detector, which may significantly enhance the maximum temperature reduction obtainable using the spin-Peltier effect, was already accomplished in measurements by \citet{Sola2019-ug}, who experimentally proved the reciprocity between spin-Seebeck and spin-Peltier effect in a YIG$|$Pt bilayer. \citet{Uchida2017-zf} achieved an enhancement of the spin-Peltier effect in ferromagnet--normal-metal multilayers upon increasing the number of layers. Their observed enhancement of the temperature modulation (integrated over all layers) with an increased number of layers can be attributed to a reduced spin backflow compared to a single bilayer of Fe$_3$O$_4|$Pt. Both Refs. \onlinecite{Sola2019-ug} and \onlinecite{Uchida2017-zf} aimed to separate the spin-Peltier effect from Joule heating by an {\it ac} modulation of the source voltage. Without such modulation, Joule heating and (nonlocal) spin-Peltier cooling are in direct competition, and one has to resort to nonlocal probes to disentangle the two effects.

\section{Conclusion}
\label{sec:discussion}

The comprehensive linear-response theory discussed in this article describes magnon-mediated transport effects in multilayers combining ferromagnetic insulators and normal metals for driving frequencies ranging from the {\it dc} limit well into the {\it ac} (THz) regime. Our approach, which describes the coupling across the magnetic-insulator--normal-metal interface using the spin-mixing conductance and otherwise relies on semi-phenomenological transport equations, 
establishes a consistent, unified description of a plethora of spintronic effects reported in the literature. These include the spin-Hall magnetoresistance \cite{Chen2013-gf,Chen2016-pc,Zhang2019-zv,Reiss2021-em}, magnon-mediated electric current drag \cite{Zhang2012-fy, Li2016-ye, Wu2016-cs}, nonlocal magnon-mediated magnetoresistance \cite{Wang2018-qd}, as well as spin-caloritronic effects \cite{Liu2021-ny,Flipse2014-kh}. Having a common theoretical framework allows a quantitative comparison between mechanisms based on incoherent and coherent magnon transport and between different driving frequencies.

In principle, in addition to the magnon-mediated effect considered here, Coulomb interactions may also directly cause an electrical current drag. In the {\it dc} limit, such Coulomb drag is usually very small in metals, because it requires an electron-hole asymmetry in the conductivity of the normal metal \cite{Narozhny2016-vl}. At finite frequency, depending on the precise measurement geometry, a Coulomb-mediated current drag can also be caused by parasitic capacitive couplings. In either case, the magnon-mediated electrical current drag can be distinguished from Coulomb-mediated effects  by its unique dependence on the magnetization direction.

The effects discussed in this article also have nonlinear counterparts, most notably the unidirectional spin-Hall magnetoresistance \cite{Avci2015-pk,Avci2018-uv,Liu2021-ny} and the spin-torque diode effect \cite{Tulapurkar2005-ew, Sankey2006-cf,Liu2011-sy,Kondou2012-kt,Ganguly2014-zw,Schreier2015-pt,Sklenar2015-gm}. The nonlinear response can not only be distinguished from the linear response by its dependence on the strength of the driving field, it also has a different characteristic dependence on the magnetization of the ferromagnetic layer. The unidirectional spin-Hall magnetoresistance includes a spin-dependent modulation of the interface conductivity that has an $m_y^3$ dependence on the magnetization in a longitudinal measurement setup. The spin-torque diode effect, on the other hand, arises from the precession of the magnetization in F and therefore shows spin-wave resonances and has an $m_y(1-m_y^2)$ magnetization dependence. These effects are accompanied by a spin-Seebeck contribution from Joule heating that scales with $m_y$. All these bilinear effects have in common that they are asymmetric under a reversal of the magnetization direction, which allows for the detection of magnetization switching in a two-terminal setup \cite{Avci2015-pk}. The linear response theory established in this work offers a suitable framework to also analyze the quadratic-in-applied-field response. We refer to our companion article \cite{Franke2025-nonlin} for further details.

\section*{Acknowledgments}
We thank D. A. Reiss, T. Kampfrath, P. Kuba\v{s}\v{c}\'{i}k, Q. Remy, and G. Lemut for stimulating discussions. This work was funded by the Deutsche Forschungsgemeinschaft (DFG, German Research Foundation) through the Collaborative Research Center SFB TRR 227 ``Ultrafast spin dynamic'' (Project-ID 328545488, project B03).

\section*{Data availability}
The data that support the findings of this article are openly available \cite{Franke2025-zenodo}.

\begin{appendix}
\renewcommand{\thefigure}{A\arabic{figure}}
\setcounter{figure}{0}

\section{Derivation of phenomenological transport equations from Boltzmann theory}
\label{sec:boltzmann}
In this Appendix, we derive expressions for the matrices $\mathcal{C}_{\rm m}$, $\mathcal{G}_{\rm m}$, and $\Sigma(\omega)$ in Eqs.~(\ref{eq:drhoms}), Eq.~(\ref{eq:diffeq_magnonheattransport}), and Eq.~(\ref{eq:ferromagnetlinresponse}), respectively, that describe incoherent magnon transport in a ferromagnet. Numerical estimates for these matrices are presented in Tab.~\ref{tab:derivedparameters}.

We express the magnon spin and energy density in terms of the magnon distribution function $n_{\vk}(z,t)$, which we linearize around the equilibrium distribution function $n^0(\omega_\mathbf{k}) = 1/(e^{\hbar \omega_\mathbf{k} / k_{\rm B} T} - 1)$,
\begin{equation}
    n_\mathbf{k}(z,t) = n^0(\omega_\mathbf{k}) + \left( - \frac{\partial n^0}{\partial \omega_{\vk}}\right) \psi_\mathbf{k}(z,t).
\end{equation}
Consistent with a description in terms of magnon spin and energy density and the corresponding currents, for the deviation from equilibrium $\psi_{\vk}(z,t)$ we use the {\em Ansatz}
\begin{align}
\begin{split}
  \psi_\mathbf{k}(z,t) =&\ 
  \frac{1}{\hbar}\mu_{\rm m}(z,t) + \frac{\omega_\mathbf{k}}{T} \Delta T_{\rm m}(z,t) \\
    &\,+ v_{\mathbf{k}z} \left[ \frac{1}{\hbar} \pi_{{\rm m}\mu}(z,t) + \frac{\omega_\mathbf{k}}{T} \pi_{{\rm m}T}(z,t) \right],
\end{split}
\label{eq:psiansatz}
\end{align}
where $\mu_{\rm m} = e u_{\rm m}$ is the local magnon chemical potential, $\Delta T_{\rm m} = (e / k_{\rm B}) u_{\rm mQ}$ the difference between the local magnon temperature and the temperature $T$ of the phonon bath (which we assume to be constant), and $\pi_{{\rm m}\mu}$ and $\pi_{{\rm m}T}$ are generalized potentials that determine the magnon spin and energy currents.

We combine the magnon spin and energy densities $(\hbar/2 e) \rho_{\rm ms}$ and $(k_{\rm B} T/2 e) \rho_{{\rm m Q}}$ into a two-component spinor,
\begin{equation}
  {\cal P}_{\rm m}(z,t) = \begin{pmatrix} \rho_{{\rm ms}}(z,t) \\ \rho_{{\rm mQ}}(z,t)
  \end{pmatrix},
\end{equation}
where we used equivalent charge units. In terms of the distribution function $n_{\vk}(z,t)$, ${\cal P}_{\rm m}(z,t)$ reads
\begin{equation}
  {\cal P}_{\rm m}(z,t) = \frac{2 e}{(2 \pi)^3}
  \int d\vk \left( - \frac{\partial n^0}{\partial \omega_{\vk}}\right)
  \psi_\mathbf{k}(z,t)
  \begin{pmatrix}
      1 \\
      \frac{\hbar \omega_{\vk}}{k_{\rm B} T}
  \end{pmatrix}.
\label{eq:rhom_density}
\end{equation}
Substitution of the {\em Ansatz} (\ref{eq:psiansatz}) then gives Eq.~\eqref{eq:drhoms} of the main text with
\begin{equation}
    {\cal C}_{\rm m} = \frac{2 e^2}{\hbar}
    \begin{pmatrix}
        F_{0,0} & F_{0,1} \\
        F_{0,1} & F_{0,2}
    \end{pmatrix},
\label{eq:calCm}
\end{equation}
where
\begin{align}
  \label{eq:Fdef}
  F_{m,n} =&\ \frac{1}{(2 \pi)^3}
  \\ \nonumber &\ \mbox{} \times
  \int d\vk\,
  \left( \frac{\hbar v_{\vk z}^{2}}{4 \Dex k_{\rm B} T} \right)^m
  \left( \frac{\hbar \omega_{\vk}}{k_{\rm B} T} \right)^n
  \left( - \frac{\partial n^0}{\partial \omega_{\vk}}\right).
\end{align}
Explicit expressions for the moments $F_{m,n}$ may be obtained for the magnon dispersion
\begin{equation}
  \omega_k = \omega_0 + \Dex k^2.
  \label{eq:dispersion}
\end{equation}
In this case, the integrations over $\vk$ in Eq.~(\ref{eq:Fdef}) can be replaced by integrations over $\omega$ and one finds
\begin{align}
\begin{split}
  F_{m,n} =&\,
  \frac{1}{2m+1}
  \sqrt{\frac{k_{\rm B} T}{\hbar \Dex^3}}
  \int_{\xi_0}^{\infty} d\xi
  \frac{\xi^n (\xi - \xi_0)^{m+1/2}}{16 \pi^2 \sinh^2(\xi/2)},
\end{split}
\end{align}
where $\xi_0 = \hbar \omega_0/k_{\rm B} T$. For $n=0$, one thus finds
\begin{equation}
  F_{m,0} =
  \frac{\Gamma(m+1/2)}{8 \pi^2 \Dex}
  \sqrt{\frac{k_{\rm B} T}{\hbar \Dex}}
  \mbox{Li}_{m+1/2}(e^{-\xi_0}),
\end{equation}
where $\mbox{Li}(z)$ is the polylogarithm and $\Gamma$ the gamma function,
whereas moments $F_{m,n}$ with $n=1$ and $n=2$ can be obtained from the recursion relations
\begin{align}
\begin{split}
  F_{m,1} =&\ \xi_0 F_{m,0} + \frac{2m+3}{2m+1} F_{m+1,0}, \\
  F_{m,2} =&\ \xi_0^2 F_{m,0} + 2\frac{2m+3}{2m+1} \xi_0 F_{m+1,0}
  \\ &\, + \frac{2m+5}{2m+1} F_{m+2,0}.
\end{split}
\end{align}
Numerical values for the coefficients of ${\cal C}_{\rm m}$, using the parameters listed in Tab. \ref{tab:materialparameters}, are given in Tab.\ \ref{tab:derivedparameters}. In the limit $\xi_0 \ll 1$, which is applicable to YIG at room temperature, one may approximate 
\begin{align}
  F_{m,n} \approx&\ \frac{\Gamma(m + n + 1/2)}{8 \pi^2 \Dex}
  \sqrt{\frac{k_{\rm B} T}{\hbar \Dex}} \\
  &\, \times 
  \left\{ \begin{array}{ll} \sqrt{\pi/\xi_0} & \mbox{if $m = n = 0$}, \\
  \frac{2m+2n+1 \vphantom{M^M_M}}{2m + 1}
  \zeta(m + n + 1/2) & \mbox{if $m +n > 0$},
  \end{array} \right.
  \nonumber
\end{align}
with $\zeta$ the Riemann zeta function.

This simple model estimate for the volumetric magnon heat capacity $[{\cal C}_{{\rm m}}]_{22}$ is  (numerically) close to the result obtained in Ref.~\onlinecite{Boona2014-cm} and agrees with experimental data at low temperatures $T < 10 \, \mathrm{K}$. At room temperature it is difficult to directly measure the magnon contribution to the heat capacity by taking the difference of high-magnetic-field and low-field heat capacities because of the large magnetic field needed to freeze out magnons \cite{Rezende2015-lz}.
Experiment \cite{Boona2014-cm} and theory \cite{Rezende2015-lz} show that the increase of heat capacity at higher temperatures is likely to be slower than $T^{3/2}$. Although this means that Eq.~(\ref{eq:calCm}) probably overestimates the magnon heat capacity at $T=300 \, \mathrm{K}$, the numerical value obtained from Eq.~(\ref{eq:calCm}) is still of the same order of magnitude as other values given in the literature, see, {\em e.g.}, Ref.~\onlinecite{Rezende2015-lz}.

For the magnon spin and heat current densities ${\cal I}(z,t)$ we find in the same manner, again using two-component spinor notation,
\begin{equation}
  {\cal I}(z,t)
  = \mathcal{V}_{\rm m}
  \Pi(z,t),
\end{equation}
where $\Pi(z,t) = (\pi_{{\rm m}\mu}(z,t) / e, k_{\rm B} \pi_{{\rm m}T}(z,t) / e)^{\rm T}$ and
\begin{align}
\begin{split}
  {\cal V}_{\rm m} =&\ 
  \frac{2 e^2}{\hbar} \int \frac{d\vk}{(2 \pi)^3} \, v_{\vk z}^2
  \left( - \frac{\partial n^0}{\partial \omega_{\vk}}\right) \\
  &\, \times \begin{pmatrix} 1 & \hbar \omega_{\vk}/k_{\rm B} T \\ \hbar \omega_{\vk}/k_{\rm B} T &
    (\hbar \omega_{\vk} /k_{\rm B} T)^2 \end{pmatrix} \\ =&\
  \frac{2 e^2}{\hbar} \frac{4 \Dex k_{\rm B} T}{\hbar}
  \begin{pmatrix} F_{1,0} & F_{1,1} \\
    F_{1,1} & F_{1,2}
  \end{pmatrix}.
\end{split}
\label{eq:calVm}
\end{align}

To find expressions for the matrices ${\cal G}_{\rm m}$ and $\Sigma_{\rm m}$, introduced in the spin continuity equation \eqref{eq:diffeq_magnonheattransport} and Ohm's law for thermal magnons in Eq.~\eqref{eq:ferromagnetlinresponse}, respectively, we use the Boltzmann equation for the linearized distribution function $\psi_\mathbf{k}(z,t)$,
\begin{align}
  \frac{\partial \psi_\mathbf{k}(z,t)}{\partial t} + v_{\vk z} \frac{\partial \psi_\mathbf{k}(z,t)}{\partial z}
  =&\ \sum_{\alpha}
  \int d\vk' \Gamma^{(\alpha)}_{\vk,\vk'} \psi_{\vk'}(z,t),
\label{eq:boltzmann}
\end{align}
where the summation is over different relaxation processes, labeled by $\alpha$, and $\Gamma^{(\alpha)}_{\vk,\vk'}$ denotes the corresponding transition rates. To obtain closed equations for $\mu_{\rm m}(z,t)$ and $\Delta T_{\rm m}(z,t)$, we multiply Eq.~(\ref{eq:boltzmann}) by $\hbar (-\partial n^0/\partial \omega_{\vk})$ and by $\hbar \omega_{\vk} (-\partial n^0/\partial \omega_{\vk})$, substitute the {\em Ansatz} (\ref{eq:psiansatz}) for the linearized distribution function, and integrate over $\vk$. Integrals on the r.h.s.\ containing odd powers of $v_{\vk z}$ vanish, so that the r.h.s.\ is a linear function of $\mu_{\rm m}(z,t)$ and $\Delta T_{\rm m}(z,t)$ only. The resulting equation has the form of Eq.~\eqref{eq:diffeq_magnonheattransport}, with
\begin{align}
  {\cal G}_{\rm m} =&\, -
  \frac{2 e^2}{\hbar} \int \frac{d\vk d\vk'}{(2 \pi)^3}
  \left( - \frac{\partial n^0}{\partial \omega_{\vk}}\right)
  \\ &\, \times
  \sum_{\alpha} \Gamma_{\vk,\vk'}^{(\alpha)}
  \begin{pmatrix} 1 & \hbar \omega_{\vk'}/k_{\rm B} T \\ \hbar \omega_{\vk}/k_{\rm B} T &
    \hbar^2 \omega_{\vk} \omega_{\vk'} /(k_{\rm B} T)^2 \end{pmatrix}. \nonumber
\end{align}

The relaxation-time approximation corresponds to the simple choice
\begin{equation}
  \Gamma_{\vk,\vk'}^{(\alpha)} = -\frac{1}{\tau_{\alpha}}
  \left[ \delta_{\vk,\vk'} - \frac{1}{(2 \pi)^3}
    \left( - \frac{\partial n^0}{\partial \omega_{\vk'}} \right)
    c_{\vk,\vk'}^{(\alpha)} \right],
  \label{eq:relaxationtime}
\end{equation}
where $\delta_{\vk,\vk'}$ is the Dirac delta function and $\tau_{\alpha}$ is the relaxation time, which is independent of $\vk$. The second term between the square brackets is a phenomenological correction term that ensures conservation of magnon spin density and/or magnon energy density, if applicable \cite{Mermin1970-kn}. We consider five relaxation processes for magnons: exchange-based spin-conserving magnon-magnon ($\alpha = \mbox{``${\rm m},{\rm ex}$''}$) and magnon-phonon scattering ($\alpha = \mbox{``${\rm mp},{\rm ex}$''}$), relativistic spin-non-conserving magnon-magnon ($\alpha = \mbox{``${\rm m},{\rm rel}$''}$) and magnon-phonon scattering ($\alpha = \mbox{``${\rm mp},{\rm rel}$''}$), and elastic magnon-impurity scattering ($\alpha = \mbox{``${\rm el}$''}$). For elastic magnon-impurity scattering, which relaxes the magnon currents, but not the magnon spin and energy densities, and for exchange-based magnon-magnon scattering one has
\begin{align}
  c_{\vk,\vk'}^{(\alpha)} =&\,
  \frac{2 e^2}{\hbar} \begin{pmatrix} 1 & \hbar \omega_{\vk}/k_{\rm B} T \end{pmatrix}
      {\cal C}_{\rm m}^{-1} \begin{pmatrix} 1 \\ \hbar \omega_{\vk'}/k_{\rm B} T \end{pmatrix},
\end{align}
with $\alpha = \mbox{``${\rm m},{\rm ex}$''}$ or $\alpha = \mbox{``${\rm el}$''}$.
(If umklapp processes were ruled out, exchange-based magnon-magnon scattering would also conserve magnon momentum density, which would result in a modified correction term.) Similarly, for exchange-based magnon-phonon scattering, which conserves magnon spin density, but not magnon energy density, and for relativistic three-magnon scattering, which conserves magnon energy density, but not the magnon spin density, one has
\begin{align}
  c_{\vk,\vk'}^{{\rm mp},{\rm ex}} =&\ \frac{2 e^2}{\hbar} \frac{1}{[{\cal C}_{\rm m}]_{11}}, \\
  c_{\vk,\vk'}^{{\rm m},{\rm rel}} =&\ \frac{2 e^2}{\hbar} \left(\frac{\hbar}{k_{\rm B} T} \right)^2 \frac{\omega_{\vk} \omega_{\vk'}}{[{\cal C}_{\rm m}]_{22}}.  
\end{align}
Finally, for relativistic magnon-phonon scattering, which conserves neither the magnon spin density nor the magnon current density, one may set
\begin{equation}
  c_{\vk,\vk'}^{{\rm mp},{\rm rel}} = 0.
\end{equation}

Calculating the four elements of the relaxation matrix ${\cal G}_{\rm m}$ in the relaxation-time approximation (\ref{eq:relaxationtime}) then results in
\begin{align}
  {\cal G}_{\rm m} =&\
  {\cal C}_{\rm m} \tau^{-1}_{{\rm mp},{\rm rel}}
  \nonumber \\ &\,
  + (1-\gamma) \begin{pmatrix}
    [{\cal C}_{\rm m}]_{11} \tau^{-1}_{{\rm m},{\rm rel}} & 0 \\ 0 &
    [{\cal C}_{\rm m}]_{22} \tau^{-1}_{{\rm mp},{\rm ex}} \end{pmatrix},
  \label{eq:calGmcalCmrelation}
\end{align}
with
\begin{equation}
    \gamma = \frac{[{\cal C}_{\rm m}]_{12} [{\cal C}_{\rm m}]_{21}}{[{\cal C}_{\rm m}]_{11} [{\cal C}_{\rm m}]_{22}},
\label{eq:gammaratio}
\end{equation}
and the total momentum relaxation time $\tau_{\rm m}$ in Eq.~\eqref{eq:capitalsigma} becomes Eq.~\eqref{eq:magnonlifetime}.

Numerical estimates for the scattering times are given in Tab.~\ref{tab:derivedparameters}.
Without the terms proportional to $\gamma$ and upon identifying (notation of Ref.~\onlinecite{Cornelissen2016-wy}) \mbox{$\Gamma_{\rm s \mu} = \tau_{{\rm mp},{\rm rel}}^{-1} + \tau_{{\rm m},{\rm rel}}^{-1}$} and \mbox{$\Gamma_{22} = \tau_{{\rm mp},{\rm rel}}^{-1} + \tau_{{\rm mp},{\rm ex}}^{-1}$} as well as the cross terms \mbox{$\Gamma_{\rm sT} = \tau_{{\rm mp},{\rm rel}}^{-1} [{\cal C}_{\rm m}]_{22}^{-1} [{\cal C}_{\rm m}]_{12}$ (in $1/\mathrm{Js}$)} and $\Gamma_{\rm Q \mu} = - \tau_{{\rm mp},{\rm rel}}^{-1} [{\cal C}_{\rm m}]_{12} [{\cal C}_{\rm m}]_{11}^{-1}$, Eqs.~(\ref{eq:diffeq_magnonheattransport}) and (\ref{eq:calGmcalCmrelation}) reproduce the spin and heat continuity equations for magnon transport from Ref.~\onlinecite{Cornelissen2016-wy}.

In a similar manner, the linear-response equation (\ref{eq:ferromagnetlinresponse}) can be obtained from the Boltzmann equation (\ref{eq:boltzmann}) upon multiplication by $(2 e / \hbar) \hbar v_{\vk z} (-\partial n^0/\partial \omega_{\vk})$ and by $(2 e / k_{\rm B} T) \hbar \omega_{\vk} v_{\vk z} (-\partial n^0/\partial \omega_{\vk})$, followed by integration over $\vk$. This gives
\begin{align}
  \frac{\partial}{\partial t}
  {\cal I}(z,t)
  + {\cal V}_{\rm m} \frac{\partial}{\partial z}
  {\cal U}_{\rm m}(z,t)
  =&\,
  - \tau_{\rm m}^{-1}
  {\cal I}(z,t)
\end{align}
where $\tau_{\rm m}$ is a $2 \times 2$ matrix defined as
\begin{align}
  \tau_{\rm m}^{-1} = \sum_{\alpha} \tau_{{\rm m}}^{(\alpha)-1},
\end{align}
with
\begin{align}
  \tau_{{\rm m}}^{(\alpha)-1} {\cal V}_{\rm m} =&\,
  - \frac{2 e^2}{\hbar} \int \frac{d\vk d\vk'}{(2 \pi)^3}
  \left( - \frac{\partial n^0}{\partial \omega_{\vk}}\right) \Gamma_{\vk,\vk'}^{(\alpha)} v_{\vk z} v_{\vk' z} \nonumber \\
&\ \ \ \ \ \mbox{} \times
  \begin{pmatrix} 1 & \hbar \omega_{\vk}/k_{\rm B} T \\ \hbar \omega_{\vk}/k_{\rm B} T &
    (\hbar \omega_{\vk}/k_{\rm B} T)^2 \end{pmatrix}.
\end{align}

Hence, Eq.~(\ref{eq:ferromagnetlinresponse}) follows, with
\begin{equation}
  \Sigma(\omega) = (\openone - i \omega \tau_{\rm m})^{-1} \tau_{\rm m} 
  {\cal V}_{\rm m},
\end{equation}
where $\openone$ is the $2 \times 2$ unit matrix. In the relaxation-time approximation (\ref{eq:relaxationtime}) one has
\begin{equation}
  \tau_{\rm m}^{-1} = \openone
  \sum_{\alpha} \tau_{\alpha}^{-1},
\end{equation}
consistent with Eq.~(\ref{eq:magnonlifetime}). For the magnon conductivity $\sigma_{\rm m}$, the spin-Seebeck coefficient $L_{\rm m}$, and the magnon thermal conductivity $\kappa_{\rm m}$, we thus find, in the relaxation-time approximation
\begin{align}
  \sigma_{\rm m} =&\ \frac{8 e^2 \tau_{\rm m} \Dex k_{\rm B} T}{\hbar^2} F_{1,0},
  \label{eq:sigmam} \\
  L_{\rm m} =&\ \frac{8e \tau_{\rm m} \Dex k_{\rm B}^2 T^2}{\hbar^2} F_{1,1},
  \label{eq:L} \\
  \kappa_{\rm m} =&\ \frac{4 \tau_{\rm m} \Dex k_{\rm B}^3 T^2}{\hbar^2} F_{1,2}.
  \label{eq:kappam}
\end{align}
These three results agree with Ref.~\onlinecite{Cornelissen2016-wy}. We may overestimate $\kappa_{\rm m}$ at room temperature since the total thermal conductivity of YIG is only two times larger than our prediction for $\kappa_{\rm m}$ \cite{Padture2005-my} and the magnon contribution is often assumed to be smaller than the phonon contribution, \textit{e.g.}, in Ref.~\onlinecite{Schreier2013-cb}. Since we adjusted magnon impurity scattering to correctly reproduce the spin conductivity, we obtain a similar $\sigma_{\rm m}$ as measured in Ref.~\onlinecite{Cornelissen2015-fh}.

\section{Spin impedances}
\label{sec:impedances}
In this Appendix, we solve the system of equations established in Sec.~\ref{sec:linearresponse} and express the dimensionless linear response coefficients $\zzeta_{{\ib\ia\perp}}(\omega)$ and $\calzzeta_{{\ib\ia\parallel}}(\omega)$ introduced in Sec.~\ref{sec:conductivities} in generalized impedances.

As a first step, we solve Eqs.~\eqref{eq:nmspinrelation}, \eqref{eq:nmheatrelation}, \eqref{eq:fnspinrelationperp}, \eqref{eq:fnrelationpara}, \eqref{eq:fispinrelationperp}, and \eqref{eq:ferromagnonheatrelation} for the spin current $\vi_{{\rm s}\ia }$ through the F$|$N$\ia$ interfaces, $\ia=1,2$, in response to a source $\delta \vu_\ib$,
\begin{align}
\begin{split}
    i_{{\rm s}\perp \ia}(\omega) &= - \sum_{\ib=1}^2  (-1)^{\ib-1} Z_{\ia\ib\perp}(\omega)^{-1} \delta u_{{\rm s}\ib \perp}(\omega), \\
    {\cal I}_\ia(\omega) &= - \sum_{\ib=1}^2  (-1)^{\ib-1} {\cal Z}_{\ia\ib\parallel}(\omega)^{-1} \delta {\cal U}_{{\rm e}\ib }(\omega).
\end{split}
\label{eq:isus}
\end{align}
The total local and nonlocal spin impedances for transverse transport read
\begin{widetext}
\begin{align}
  Z_{11\perp}(\omega) =&\ Z_{{\rm N}1} + Z_{{\rm FN}1\perp}(\omega)
  + \left[\cos(k(\omega) d_{\rm F}) (Z_{{\rm N}2} + Z_{{\rm FN}2 \perp})
  -i \sin(k(\omega) d_{\rm F}) Z_{{\rm F}\perp}^{\infty}(\omega) \right]
  \nonumber \\ &\, \ \ \ \ \mbox{} \times 
  \left[ \cos(k(\omega) d_{\rm F}) Z_{{\rm F}\perp}^{\infty}(\omega) - i \sin(k(\omega) d_{\rm F})
  (Z_{{\rm N}2} + Z_{{\rm FN}2 \perp}) \right]^{-1}  Z_{{\rm F}\perp}^{\infty}(\omega),
  \label{eq:Z11perp}
  \\
  Z_{21\perp}(\omega) =&\
  \left(Z_{{\rm N}1} + Z_{{\rm FN}1\perp} \right) \left(Z_{{\rm F}\perp}^{\infty}(\omega) \right)^{-1}
  \left[ \cos(k(\omega) d_{\rm F}) Z_{{\rm F}\perp}^{\infty}(\omega)
    - i \sin(k(\omega) d_{\rm F}) (Z_{{\rm N}2} + Z_{{\rm FN}2 \perp}) \right] 
  \nonumber \\ &\, \mbox{}
  + \cos(k(\omega) d_{\rm F}) (Z_{{\rm N}2} + Z_{{\rm FN}2 \perp}) - i \sin(k(\omega) d_{\rm F}) Z_{{\rm F}\perp}^{\infty}(\omega).
  \label{eq:Z21perp}
\end{align}
The effective impedances $Z_{12\perp}(\omega) = Z_{21\perp}(\omega)$ and $Z_{22\perp}(\omega)$ are obtained from Eqs.~(\ref{eq:Z11perp}) and (\ref{eq:Z21perp}) by interchanging the indices $1 \leftrightarrow 2$.
Note that \mbox{$Z_{\ia\ia\perp}(\omega \to 0) = Z_{{\rm N}\ia } + Z_{{\rm FN}\ia \perp}$} since \mbox{$Z_{{\rm F}\perp}^{\infty}(0) = 0$}. 

The longitudinal response has generalized impedances
\begin{align}
  \mathcal{Z}_{11\parallel}(\omega) =&\
  \mathcal{Z}_{{\rm N}1} + \mathcal{Z}_{{\rm FN}1\parallel}
  \, + \left[ \sinh(\Lambda(\omega)^{-1} d_{\rm F}) 
  \mathcal{Z}_{{\rm F}\parallel}^{\infty}(\omega)
    + \cosh(\Lambda(\omega)^{-1} d_{\rm F})
    (\mathcal{Z}_{{\rm N}2} + \mathcal{Z}_{{\rm FN}2\parallel})
  \right] \nonumber \\ &\, \ \ \ \ \mbox{} \times
   \left[\cosh(\Lambda(\omega)^{-1} d_{\rm F}) 
    \mathcal{Z}_{{\rm F}\parallel}^{\infty}(\omega)
     + \sinh(\Lambda(\omega)^{-1} d_{\rm F})
     (\mathcal{Z}_{{\rm N}2} + \mathcal{Z}_{{\rm FN}2\parallel})
   \right]^{-1} \mathcal{Z}_{{\rm F}\parallel}^{\infty}(\omega),
\label{eq:calZ11para} \\
  \mathcal{Z}_{21\parallel}(\omega) =&\,
  (\mathcal{Z}_{{\rm N}1} + \mathcal{Z}_{{\rm FN}1\parallel})
  (\mathcal{Z}_{{\rm F}\parallel}^{\infty}(\omega))^{-1}
  \left[\cosh(\Lambda(\omega)^{-1} d_{\rm F}) 
  \mathcal{Z}_{{\rm F}\parallel}^{\infty}(\omega)
    + \sinh(\Lambda(\omega)^{-1} d_{\rm F})
    (\mathcal{Z}_{{\rm N}2} + \mathcal{Z}_{{\rm FN}2\parallel})
    \right] 
  \nonumber \\ &\, \mbox{}
   + \sinh(\Lambda(\omega)^{-1} d_{\rm F}) 
    \mathcal{Z}_{{\rm F}\parallel}^{\infty}(\omega)
   + \cosh(\Lambda(\omega)^{-1} d_{\rm F}) 
  (\mathcal{Z}_{{\rm N}2} + \mathcal{Z}_{{\rm FN}2\parallel}).
   \label{eq:calZ21para}
\end{align}
\end{widetext}
The effective impedances $\mathcal{Z}_{12\parallel}(\omega)$ and $\mathcal{Z}_{22\parallel}(\omega)$ are obtained from Eqs.~(\ref{eq:calZ11para}) and (\ref{eq:calZ21para}) by interchanging the indices $1 \leftrightarrow 2$.

The coefficients $\zzeta_{{\ib\ia\perp}}(\omega)$ and $\calzzeta_{{\ib\ia\parallel}}(\omega)$ defined in Eqs.~\eqref{eq:linresponse_perp} and \eqref{eq:linresponse_para}, respectively, are then readily obtained by inserting Eq.~\eqref{eq:isus} into Eq.~\eqref{eq:spincontinuity}. We find
\begin{align}
  \zzeta_{{\ib\ia\perp}}(\omega) =&\ \delta_{\ib\ia} - (-1)^{\ia+\ib} \frac{Z_{{\rm N}\ib }}{Z_{\ib\ia\perp}(\omega)},
  \label{eq:zetaperp}
\end{align}
and the $2 \times 2$ matrix
\begin{align}
  \label{eq:zetapara}
  \calzzeta_{{\ib\ia}}(\omega) =&\,
  \mathbb{I}_2 \delta_{\ib\ia} - (-1)^{\ia+\ib} {\cal Z}_{{\rm N}\ib }(\omega) \mathcal{Z}_{\ib\ia\parallel}^{-1} (\omega).
\end{align}

\section{Analytical estimates for spin impedances}
\label{sec:analytical}

\begin{figure}
    \centering
    \includegraphics[width=0.5\textwidth]{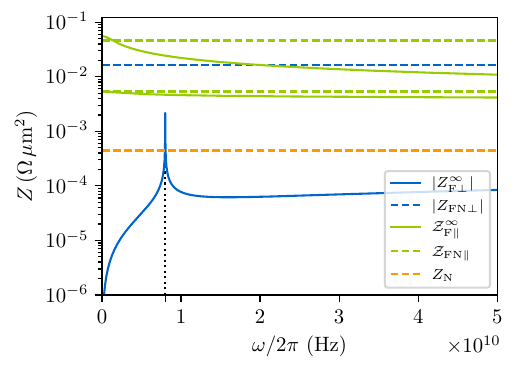}
    \caption{Spin impedances defined in Eqs.~\eqref{eq:zn}, \eqref{eq:zfnperp}, \eqref{eq:calzfnpara}, \eqref{eq:zfperpinf}, and \eqref{eq:zfparainf} for a Pt$|$YIG$|$Pt trilayer. For the matrix impedances ${\cal Z}_{{\rm FN}\parallel}$ and ${\cal Z}_{{\rm F}\parallel}^{\infty}$, this figure shows the absolute values of the two eigenvalues. The interface impedances dominate the spin transport except at low frequency, where a magnon chemical potential builds up in F, which results in a large longitudinal magnon impedance. Material and device parameters are taken from Tabs.\ \ref{tab:materialparameters} and \ref{tab:derivedparameters}.}
    \label{fig:spin_impedances}
\end{figure}

The dimensionless conductivity corrections $s_{\ib\ia}(\omega)$ and $s_{\ib\ia}''(\omega)$ are determined by the transverse and longitudinal effective impedances $Z_{\ib\ia\perp}(\omega)$ and ${\cal Z}_{\ib\ia\parallel}(\omega)$, see Eq.~(\ref{eq:sji}), whereas the conductivity correction $s_{\ib\ia}'(\omega)$ has a longitudinal contribution only. To gain a qualitative understanding of their frequency dependence, we now discuss the transverse and longitudinal effective impedances $Z_{\ib\ia\perp}(\omega)$ and ${\cal Z}_{\ib\ia\parallel}(\omega)$ separately.

For the transverse response, the effective impedance $Z_{\ib\ia\perp}(\omega)$ is dominated by the impedances of the two ferromagnet--normal-metal interfaces, except in the vicinity of resonance frequencies, see Eq.~\eqref{eq:resonancefrequencies}, where one may approximate
\begin{align}
  Z_{11\perp}(\omega) \approx&\ Z_{{\rm FN}1\perp} 
  \label{eq:resonanceapproxZperp}
  \\  &\ \mbox{}
  + \frac{Z_{{\rm FN}2\perp}}{1
    - i (-1)^{n} (\omega - \omega_n) \frac{2 e^2 d_{\rm F} s}{\hbar \omega_n (1 + \delta_{n,0})} Z_{{\rm FN}2\perp}}, \nonumber \\
  Z_{21\perp}(\omega) \approx&\  Z_{{\rm FN}1\perp} + Z_{{\rm FN}2\perp} \nonumber \\ &\ \mbox{}
  -i (-1)^{n} (\omega-\omega_n) \frac{2 e^2 d_{\rm F} s Z_{{\rm FN}1\perp} Z_{{\rm FN}2\perp}}{\hbar  \omega_n (1 + \delta_{n,0})}. \nonumber
\end{align}
(These expressions neglect the decay of coherent magnons, which is described by the Gilbert damping constant $\alpha$ in Eq.~(\ref{eq:k}).)
As a result, the transverse contributions to the local and nonlocal response coefficients $s_{\ib\ia}(\omega)$ and $s_{\ib\ia}''(\omega)$ have sharp resonant features in the vicinity of the resonance frequencies, where the sign of the resonant feature alternates between resonances. At frequency $\omega = 0$, the transverse contribution to the nonlocal conductivities is strictly zero.

If the frequency $\omega$ is large enough that $l_{m}(\omega) \ll d_{\rm F}$, $m=1,2$, the effective impedances $\mathcal{Z}_{\ia\ib\parallel}(\omega)$ become
\begin{align}
  \mathcal{Z}_{11\parallel} (\omega)
  &\approx \mathcal{Z}_{{\rm FN}1 \parallel} +
  \frac{e^{i \pi/4}}{\sqrt{\omega}} {\cal C}_{\rm m}^{-1/2} \Sigma_{\rm m}^{-1/2},
  \label{eq:largefreqapproxcalZpara} \\  \nonumber
  \mathcal{Z}_{21\parallel} (\omega) &\approx
  \frac{e^{-i \pi/4} \sqrt{\omega}}{2}
  \mathcal{Z}_{{\rm FN}1\parallel}
  \Sigma_{\rm m}^{1/2} {\cal C}_{\rm m}^{1/2}
  e^{\Lambda(\omega)^{-1} d_{\rm F}}
  \mathcal{Z}_{{\rm FN}2\parallel},
\end{align}
where $\mathcal{Z}_{12\parallel}(\omega)$ and $\mathcal{Z}_{22\parallel}(\omega)$ can be obtained by interchanging the labels $1 \leftrightarrow 2$ and $\mathcal{Z}_{\ia\ia\parallel} (\omega) = \mathcal{Z}_{\ia\ia\parallel}(-\omega)^*$. The exponential increase of ${\cal Z}_{21\parallel}(\omega)$ with $d_{\rm F}$ signals an exponential suppression of the longitudinal contribution to the nonlocal response with $\sqrt{\omega}$ in the limit of large frequency $\omega$. For the material parameters of a Pt$|$YIG$|$Pt trilayer with $d_{\rm F}$ in the nm range, this means that the longitudinal contribution to the nonlocal conductivities becomes vanishingly small for $\omega$ in the GHz range and above.

In the limit of small thickness $d_{\rm F} \ll l_{\rm T} (\ll l_{\mu})$ of the ferromagnetic layer, the longitudinal impedances become
\begin{align}
\begin{split}
  {\cal Z}_{11\parallel}(0) \approx &\ {\cal Z}_{21\parallel}(0) \\ \approx &\ \mathcal{Z}_{{\rm FN}1 \parallel} + \mathcal{Z}_{{\rm FN}2 \parallel} + d_{\rm F} \Sigma_{\rm m}^{-1}(0).
\end{split}
\label{eq:Zdsmall}
\end{align}
If $l_T \ll d_{\rm F} \ll l_{\mu}$, no simple approximate expressions for the effective impedances can be derived. Since in this intermediate regime a nonequilibrium magnon distribution exists in the entire ferromagnetic layer, the order of magnitude of the nonlocal response is still that of Eq.~(\ref{eq:Zdsmall}), albeit with a different numerical prefactor. In the limit of large $d_{\rm F}$, such that $(l_{\rm T} \ll ) l_{\mu} \ll d_{\rm F}$, one finds 
\begin{align}
\begin{split}
  \mathcal{Z}_{11\parallel} (0)
  \approx &\ \mathcal{Z}_{{\rm FN}1 \parallel} +
  {\cal Z}_{{\rm F}\parallel}^{\infty}(0) \\ 
  \mathcal{Z}_{21\parallel} (0) \approx &\,
  \frac{1}{2}
  (\mathcal{Z}_{{\rm FN}1\parallel} {\cal Z}_{{\rm F}\parallel}^{\infty}(0)^{-1}
    + \openone)
    e^{\Lambda(0)^{-1} d_{\rm F}} \\ &\ \mbox{} \times
  (\mathcal{Z}_{{\rm FN}2\parallel} + {\cal Z}_{{\rm F}\parallel}^{\infty}(0)),
\end{split}
\label{eq:lowfreqapproxcalZpara}
\end{align}
where
\begin{equation}
  {\cal Z}_{{\rm F}\parallel}^{\infty}(0) = {\cal G}_{\rm m}^{-1/2} \Sigma_{\rm m}^{-1/2}
\end{equation}
is the zero-frequency spin impedance of the ferromagnet. For $\mathcal{Z}_{21\parallel} (0)$, this gives the order-of-magnitude estimate
\begin{align}
\begin{split}
  \mathcal{Z}_{21\parallel} (0)
  \sim &\ \frac{2e^2 k_{\rm B} T \sqrt{\tau_{\rm m}}}{\hbar^2 \Dex \sqrt{\tau_{\rm rel}}}
  \left(Z_{{\rm FN}1\parallel}  + 
   \frac{\hbar^2 \Dex \sqrt{\tau_{\rm rel}}}{2e^2 k_{\rm B} T \sqrt{\tau_{\rm m}}}\right)
\\
  &\ \times 
   \left(Z_{{\rm FN}2\parallel}  + 
    \frac{\hbar^2 \Dex \sqrt{\tau_{\rm rel}}}{2e^2 k_{\rm B} T \sqrt{\tau_{\rm m}}}\right)
    e^{d_{\rm F}/l_{\mu}},
\end{split}
\end{align}
from which one easily derives an order-of-magnitude estimate for the nonlocal conductivities.  

In Fig.\ \ref{fig:spin_impedances} we show (the absolute value of) the transverse impedances $Z_{{\rm FN}\perp}$ and $Z_{{\rm F}\perp}^{\infty}$ as a function of frequency, as well as (the absolute values of) the eigenvalues of the longitudinal impedances ${\cal Z}_{{\rm N}}$, ${\cal Z}_{{\rm FN}\parallel}$ and ${\cal Z}_{{\rm F}\parallel}^{\infty}$, using the parameters from Tabs.\ \ref{tab:materialparameters} and \ref{tab:derivedparameters}. 

\end{appendix}

\typeout{}
\bibliography{main.bib}

\end{document}